\documentclass{aa}  
\usepackage{xcolor}
\usepackage{graphicx}
\usepackage{txfonts}
\usepackage{newtxtext,newtxmath}

\usepackage[T1]{fontenc}
\usepackage{ae,aecompl}
\usepackage{natbib}

\usepackage{amsmath}	
\usepackage[normalem]{ulem}
\useunder{\uline}{\ul}{}
\usepackage{placeins}
\usepackage{rotating}
\usepackage{float}
\usepackage{threeparttable}
\usepackage[rightcaption]{sidecap}

\begin{document}

   \title{Spectroscopic metallicities and first $\alpha$-element abundances of RR Lyrae stars in Baade's Window}

   \author{J. Olivares Carvajal
          \inst{\textbf{1},2}
          \and
          Á. Rojas-Arriagada\inst{3,4}
          \and
          M. Zoccali\inst{2}
          \and
          B. Acosta-Tripailao\inst{2}
          \and
          M. De Leo\inst{5,6}
          \and
          R. Albarracín\inst{7,8}
          \and
          M. Sanchez-Benavente\inst{9,10}
          \and
          E. Valenti\inst{11,12}
          \and
          G. Bono\inst{13,14}
          \and
          S. Duffau\inst{15}
          \and
          L. Sbordone\inst{16}
          \and
          M. Fabrizio\inst{14,17}
          \and
          V.F. Braga\inst{14}
          }

   \institute{INAF - Osservatorio Astronomico di Capodimonte, Salita Moiariello 16, 80131, Naples, Italy\\
              \email{julio.olivarescarvajal@inaf.it}
         \and
             Instituto de Astrof\'isica, Pontificia Universidad Cat\'olica de Chile, Av. Vicu\~na Mackenna 4860, 782-0436 Macul, Santiago, Chile
         \and
             Departamento de F\'isica, Universidad de Santiago de Chile, Av. Victor Jara 3659, Santiago, Chile
         \and
             Center for Interdisciplinary Research in Astrophysics and Space Exploration (CIRAS), Universidad de Santiago de Chile, Santiago, Chile
        \and
            Dipartimento di Fisica e Astronomia, Università degli Studi di Bologna, Via Piero Gobetti 93/2, Bologna, 40129, Italy
        \and
            Osservatorio di Astrofisica e Scienza dello Spazio di Bologna, INAF, Via Piero Gobetti 93/3, Bologna, 40129, Italy
        \and
             Max Planck Institute for Astronomy, D-69117 Heidelberg, Germany
        \and
            Fakultät für Physik und Astronomie, Universität Heidelberg, Im Neuenheimer Feld 226, 69120 Heidelberg, Germany
        \and
            IAC – Instituto de Astrofísica de Canarias, calle Via Láctea s/n, E-38205 La Laguna, Tenerife, Spain
        \and
            Departamento de Astrofísica, Universidad de La Laguna, E-38206 La Laguna, Tenerife, Spain
        \and 
            European Southern Observatory, Karl Schwarzschild-Straße 2, 85748 Garching bei München, Germany
        \and
            Excellence Cluster ORIGINS, Boltzmann–Straße 2, D–85748 Garching bei München, Germany
        \and
            Department of Physics, Università di Roma Tor Vergata, via della Ricerca Scientifica 1, Roma, 00133, Italy
        \and
            INAF - Osservatorio Astronomico di Roma, via Frascati 33, Monte Porzio Catone, 00078, Italy
        \and
            AUI/NRAO - National Radio Astronomy Observatory, Associated Universities, Inc., Av. Nueva Costanera 4091, Santiago, Chile
        \and 
            ESO – European Southern Observatory, Alonso de Cordova 3107, Vitacura, Santiago, Chile
        \and
            ASI - Space Science Data Center, Via del Politecnico s.n.c., I-00133 Roma, Italy
        }

   \date{Accepted May 27, 2026. Received December 29, 2025}

 
  \abstract
   {The shape and kinematics of the metal-poor stellar component of the Galactic bulge are still poorly characterized, and therefore, the origin of this component is not yet strongly constrained. RR Lyrae stars in the bulge have been reported to be associated with the spheroidal, relatively metal-poor component. They offer a way to trace this component with precise distances and therefore the possibility to calculate orbits and minimize contaminations.
   While a few studies of RR Lyrae spectra with medium/high resolution are now available, none of them target stars in the
   Galactic bulge.}
   {We present here a spectroscopic determination of Fe and $\alpha$-element abundances for RR Lyrae stars in the Galactic bulge, with
    the main goal of providing a benchmark to calibrate other metallicity indicators, appropriate for this specific stellar
    population.}
   {We analyzed FLAMES/GIRAFEE spectra of 78 RR Lyrae stars (60 ab-type and 18 c-type). We applied a full-spectrum fitting technique to obtain the spectroscopic metallicity and overall $\alpha$-element abundance. Distances are derived by means of a period-luminosity-metallicity relation, and orbits are computed by combining the radial velocities derived here with the proper motions from \textit{Gaia} DR3.}
   {The resulting metallicities peak at $\rm [Fe/H]_{median} = -1.34 \pm 0.04$ and $-1.44 \pm 0.08$ dex for ab and c-types respectively. The majority of the bulge RR Lyrae are metal-poor stars with relatively high $\alpha$-element abundances around $\rm [\alpha/Fe]\sim0.25 \pm 0.03$ dex. We used our spectroscopic measurements to test different methods for deriving metallicities based on photometry, which utilize Fourier parameters in the light curves of the RR Lyrae. The data suggest a possible correlation between the metallicity difference and the [$\alpha$/Fe] ratio, which needs to be investigated further. There are some ab-type RR Lyrae that show metallicities higher than -1 dex and low $\rm [\alpha/Fe]$ values. We studied these stars kinematically and found a difference between three stars with similar [$\alpha$/Fe] values and the main group, indicating that they may be slightly younger and correspond to the disk population. }
   {}


   \maketitle
%

\section{Introduction}

RR Lyrae (RRL) stars are pulsating variable stars extensively used to trace ancient stellar populations. These stars have ages exceeding 10 Gyr, as their progenitors are low-mass stars, roughly 0.6 to 0.8$M_\odot$ in the helium-core burning phase. In the context of our Galaxy, they are frequently observed in halo globular clusters \citep{Bhardwaj2022,CruzReyes2024}, the bulge \citep{soszynski19, Clementini2023, Zoccali2024}, and, more recently, in the disk \citep{Olivares2024, DOrazi2024}. The Galactic bulge, in particular, is of significant interest for understanding the formation and evolution of the Milky Way, as it constitutes a predominantly old component, accounting for at least 25\% of the total stellar mass within a highly dense region \citep[e.g.,][]{Cao2013, Valenti2016, portail2017, Simion2017}.\\ 
RRL stars exhibit periods ranging from 0.2 to 1 day and display recognizable light curves, which are valuable for determining mean magnitudes. These stars adhere to a tight period-luminosity-metallicity (PLZ) relation in the $\rm K_s$-band \citep[see e.g.,][for recent calibrations]{Neeley2019,Bhardwaj2024,Zgirski2023,Prudil2024a}. This PLZ relation is particularly useful for estimating precise distances within the Galactic context, achieving errors of around 5\%. However, a notable concern regarding this relation is the dependence on metallicity, which is often derived from photometric data. Such metallicity estimates can introduce biases in the calculated distances. Consequently, it is crucial to understand how photometric metallicities and their associated uncertainties are determined to correctly propagate the errors in the resulting distance measurements.\\
Recent studies have demonstrated that photometric metallicities can be determined from the light curves of RRL variables \citep{Mullen2021,Iorio2021,Dekany2022,Clementini2023,Li2023,Jurcsik2023}. These methods analyze the shape of the light curves and the associated Fourier parameters to evaluate how metallicity affects that shape. The method is advantageous since it is more efficient to obtain good-quality light curves than high-resolution spectra for a stellar sample. Consequently, it is possible to obtain metallicities for a massive number of stars, larger than any spectroscopic sample; however, with large individual errors. Nevertheless, it is essential to recognize that biases are linked to the calibration of the method. For instance, it is important to ensure that the data used to calibrate the method are exactly in the same passbands as the data to be analyzed.\\ 
On the other hand, several studies have derived metallicity using spectroscopy. One caveat is that their pulsating nature affects the absorption lines in the spectrum only around the steep rising branch of fundamental pulsators, while the rest of the time remains constant \citep{Pancino2015,Magurno2018}. Consequently, knowing the pulsation phase of the spectra is relevant. Additionally, RRLs are relatively hot and generally metal-poor stars, resulting in limited absorption lines available for determining metallicities and other abundances. The $\Delta S$ method utilizes the Balmer series of hydrogen lines in conjunction with the Ca II K line to estimate metallicities \citep{Preston1959,Freeman1975,Walker1991,Suntzeff1991,Layden1994}. Recent research applying this method has significantly improved the accuracy of the metallicity estimates and reduced associated errors \citep{Chadid2017,Sneden2017,Crestani2021,Crestani2021b}. More recently, \cite{Kunder2024} demonstrated that even with intermediate to low-resolution spectra and using the calcium triplet (CaT), it is possible to obtain reliable spectroscopic metallicities that agree with findings from other studies.\\ 
\begin{figure}
	\includegraphics[width=10cm]{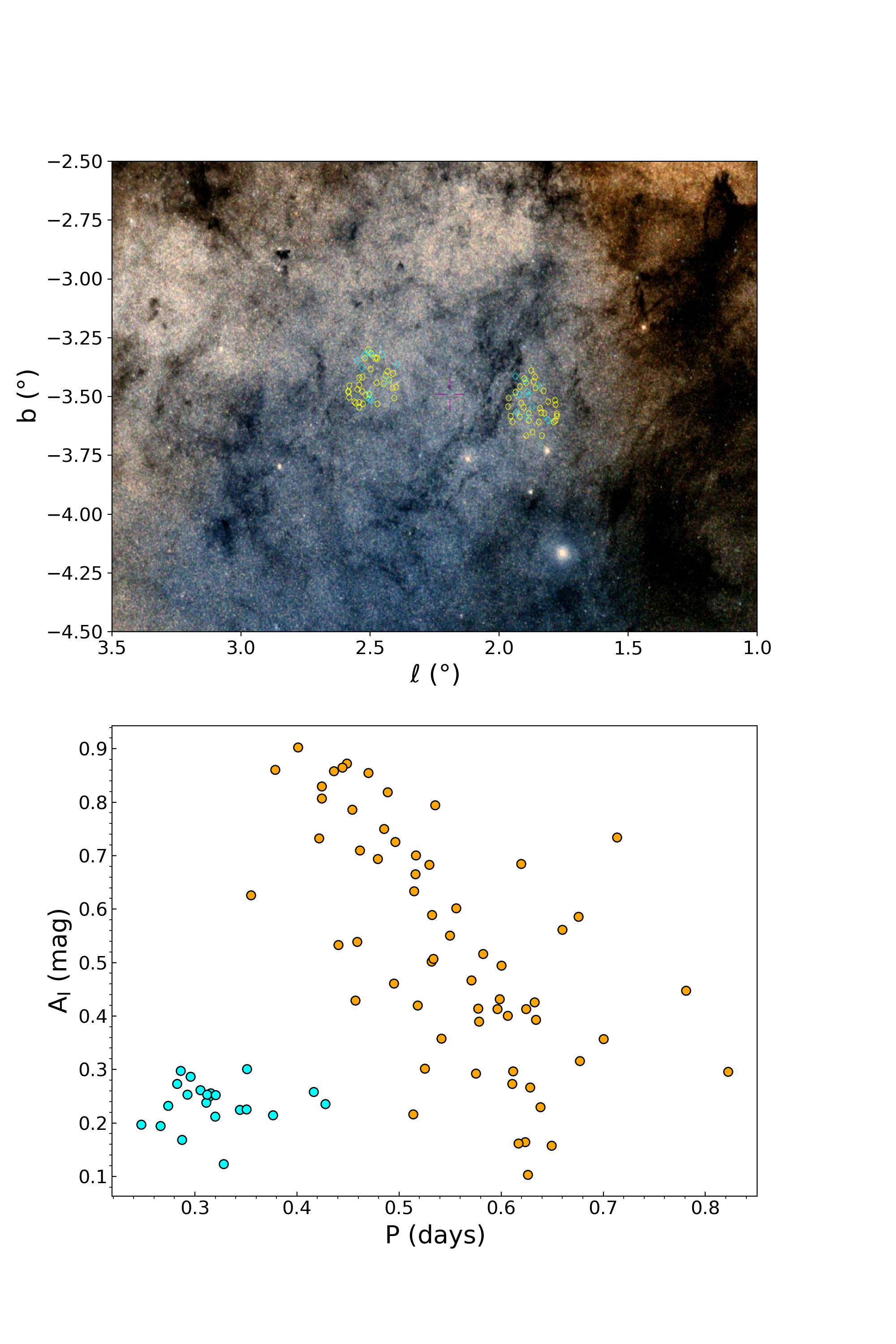}
    \caption{The GIRAFFE RRL initial sample in the Baade's Window used for this study. \textit{Top:} The location of the RRL stars in the bulge region. RRab (yellow circles) and RRc (cyan diamonds) are overlaid on an image from Aladin using a red SDSS-2 map. \textit{Bottom:} The amplitude vs period (Bailey) diagram for the RRL stars. The periods and I-band amplitudes were obtained from the OGLE-IV catalog. RRab are the orange circles, and RRc are the cyan circles.}
    \label{fig:location}
\end{figure}
The research presented in \cite{DOrazi2024}, which utilizes high-resolution spectra from the GALAH survey, establishes a novel method for accurately determining spectroscopic metallicities and abundances for both the halo and the disk through full spectral fitting of RRL stars in the solar vicinity. Additionally, the study provides measurements of $\alpha$-element abundances as well as Y and Ba abundances. These findings contribute to confirming the existence of a metal-rich tail of RR Lyrae stars, which is associated with the origins of the old disk.\\ 
The current studies indicate that new techniques have successfully achieved more precise spectroscopic measurements of metallicities and other abundances in RRL stars. Furthermore, we can test the relationship between photometric and spectroscopic metallicities with larger statistics, which will enhance the calibration of the more abundant photometric metallicity techniques. Despite these advancements, there remains a lack of research on the spectroscopic metallicities of bulge RRL stars. The study by \cite{Walker1991} is currently the only spectroscopic analysis conducted in the bulge, and to date, no research in this region has provided $\alpha$-element abundances.\\ 
We present here measurements of spectroscopic metallicities and $\alpha$-element abundances of RRL stars in two fields observed with the FLAMES/GIRAFFE spectrograph in Baade's Window, a field of the bulge well-known for its very low extinction. The paper structure is as follows: Section 2 describes the data: the spectra, light curves, distances, proper motions, and radial velocities. Section 3 presents the full spectral synthesis method employed to derive the fundamental parameters. Section 4 describes the comparison between photometric and spectroscopic metallicities. In Section 5, we explain the orbital integration employed. Section 6 shows the chemodynamical analysis of the stars. Finally,
In Section 7, we present the discussion and conclusions.
 
\begin{figure*}
	\includegraphics[width=\linewidth]{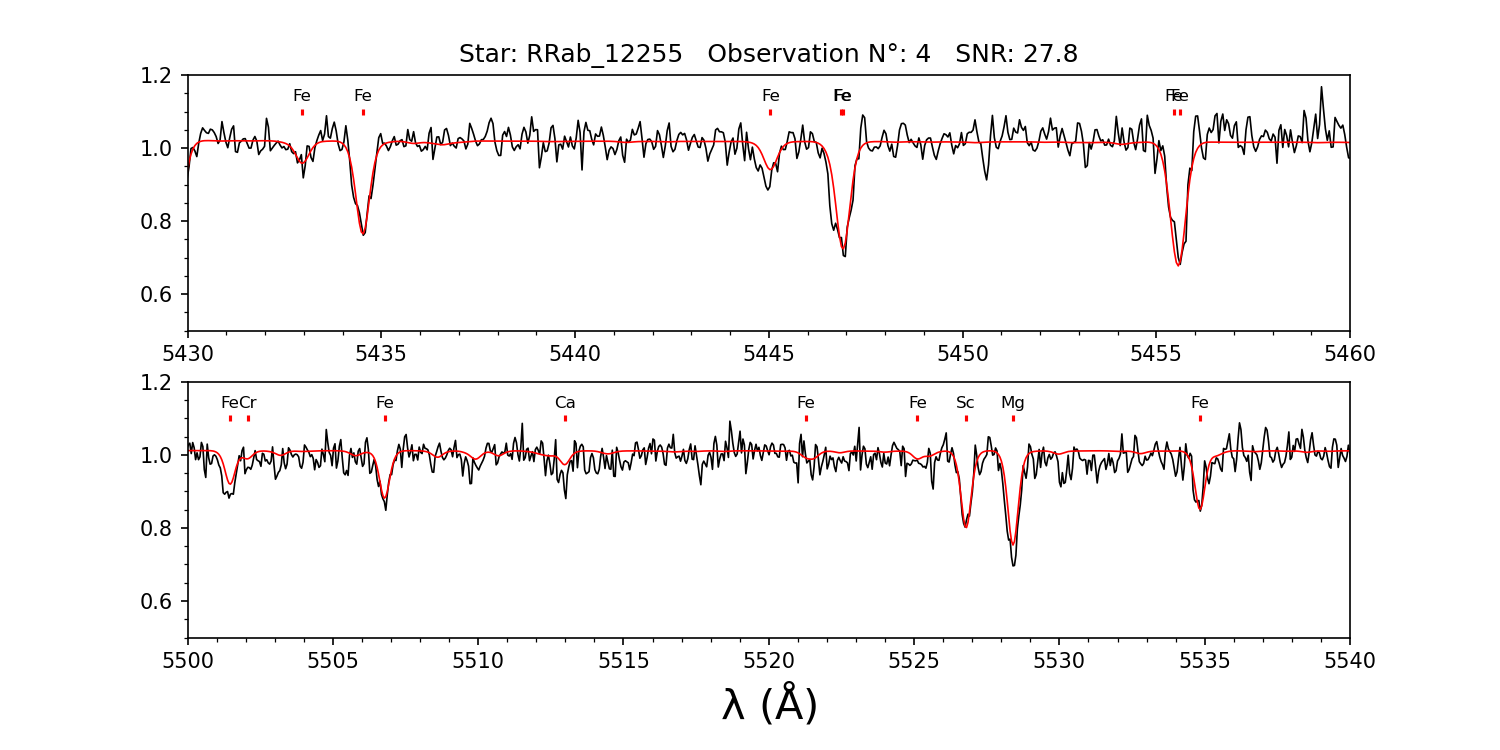}
    \caption{One example of the spectrum of the star RRab-12255 and the best fit adjusted by FERRE. The normalized and corrected observed spectrum is in black, while the best fit from the grid selected by FERRE is in red. Red ticks with labels indicate the strong lines available in this region. On top, we show a region predominantly composed of iron. In the bottom panel, the strong magnesium line (associated with the $\alpha$-element abundance) is shown to check the fitting.}
    \label{fig:spec2}
\end{figure*}
\section{Data}
\label{sec:data}

\subsection{GIRAFFE spectra}
\label{subsec:giraffe}

The spectra for this study were obtained through observations conducted with the GIRAFFE/FLAMES spectrograph \citep{GIRAFFE} installed on the 8.2-meter UT2 VLT@ESO telescope (Program ID: 093.B-0473, PI: M. Catelan). Fig.~\ref{fig:location} top panel illustrates the positions of the stars across Baade's Window in the bulge region, located at coordinates $[\ell,b]=[1.02^{\circ},-3.92^{\circ}]$. Baade's Window is particularly well-suited for optical observations due to its relatively low extinction (both absolute and differential) compared to other regions of the bulge. The sample consists of two pointings using FLAMES. The RRL stars were selected from the OGLE-III survey \citep{ogleIII}. The observation plan involved five repeated measurements of the same field fiber configuration, with exposure times of 42 min. These observations were carried out using the High-Resolution Grating 10 (HR10) within the optical range of 5339 to 5619 \AA , achieving a resolution of approximately $\rm R\sim 21500$. 

We initially had spectra for 87 RRL stars; from those, 65 were fundamental-mode RRL (RRab) and 22 were first-overtone RRL (RRc). The reduced data were obtained from the ESO phase 3 stream data release\footnote{https://www.eso.org/sci/publications/announcements/sciann17187.html}. The spectra were normalized using the \texttt{astropy} \citep{pyastronomy} package from \texttt{Python} \citep{ipython}.

The mean signal-to-noise ratio (SNR) of the individual spectra was $\rm SNR\sim 25$. Out of the 65 RRab stars analyzed, only 4 did not meet the quality criterion of SNR > 17, which was established to ensure a robust spectroscopic analysis. Below this value, the spectrum is completely dominated by the noise, and no absorption lines can be observed. In contrast, all RRc stars met this quality selection. Following the application of this criterion, we retained a total of 61 RRab stars and 22 RRc stars, each with five available spectra, except for one RRab star that had four visits, resulting in a total of 83 stars.

\subsection{Light curves}
\label{subsec:ogle}

Light curves were obtained from the OGLE-IV catalog \citep{ogleIV}. We successfully recovered all the stars. For these, we have collected I and V-band light curves. Relevant parameters derived from the I-band light curve, such as the period $P$ and the amplitude in the I-band $A_I$, are also available in the OGLE-IV catalog. The bottom panel of Fig.~\ref{fig:location} presents the Bailey diagram of our RRL sample, clearly distinguishing between RRab and RRc stars based on the parameters from the OGLE-IV catalog.

Additionally, we obtained near-infrared (near-IR) light curves for our stars from the VVV survey \citep{minniti10}, which utilized the VISTA 4.1m telescope at the ESO Paranal Observatory. This survey, equipped with the VIRCAM near-IR camera, observed the Galactic bulge and disc over a span of more than nine years. We extracted the J and $K_{_s}$-band light curves for most of our targets using point spread function (PSF) photometry, as described in \cite{Contreras2018}. The $K_{_s}$-band light curves have an average of 86 points, with those from the inner regions containing more data, approaching 120 observations. The J-band light curves typically have considerably fewer points, typically ranging from 3 to 5 points.

\begin{figure}
	\includegraphics[width=\linewidth]{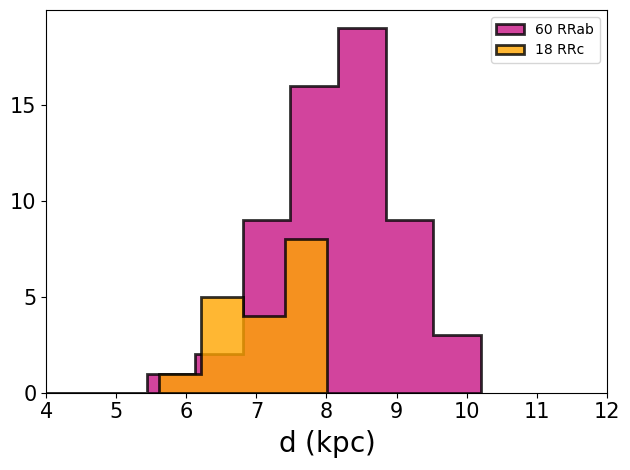}
    \caption{Histograms of the RRL distances. The red-violet distribution is for RRab stars, and the orange is for RRc stars.}
    \label{fig:dist}
\end{figure}

\subsection{Distances}
\label{subsec:dist}
To determine the distances, we used the classic distance equation
\begin{equation}
    m_\lambda-M_{\lambda}-A_{\lambda}= 5 \log{d}-5,
\end{equation}
where $m_\lambda$ is the mean magnitude, $M_\lambda$ is the absolute magnitude, and $A_\lambda$ is the extinction coefficient. The absolute magnitude was calculated using the period-luminosity-metallicity (PLZ) relation by \citet{Prudil2024a} for the $I$, $J$ and $K_s$-band.   
\begin{equation}
   M_\lambda= \alpha \log{(P)}+\beta {\rm [Fe/H]}+\gamma,
\end{equation}
where $P$ is the period of the RRL star, and [Fe/H] is the metallicity, that we derive in Sec.~\ref{sec:metalpha}. We also used the PLZ to calculate the color excess in the same manner as \cite{Prudil2025} in eq. (5) and (7). Then, the extinction coefficient can be calculated as
\begin{equation}
    A_{\lambda}=R_\lambda\times E(\lambda_1-\lambda_2),
\end{equation}
where $R_{\lambda}$ is the extinction ratio, and we adopted the values of Table 3 from \cite{Prudil2025}. We derived absolute magnitudes for each passband using the corresponding PLZ and, subsequently, this subtraction is the obtained reddening. The reddening values obtained and the extinction law from \cite{Prudil2025} are in good agreement with previous reddening maps and with the ratio $R_V\sim2.5$ previously reported in Baade's Window \citep{Nataf2013,Saha2019}. The observations were designed to avoid large changes in the extinction with values closer to $A_V\sim1.0$ mag. Finally, using this reddening as well, we obtained the distance for the RRL stars.

In order to estimate errors, we used the following formula, derived from the distance modulus expression:
\begin{equation}
    (\Delta \mathrm{d})^2= (0.46 \mathrm{d})^2 [\delta m_\lambda^2+\delta A_{\lambda}^2 + \delta M_\lambda^2],
\end{equation}
where this is the statistical error that we considered for our distance.

\begin{figure*}
\centering
	\includegraphics[width=12cm]{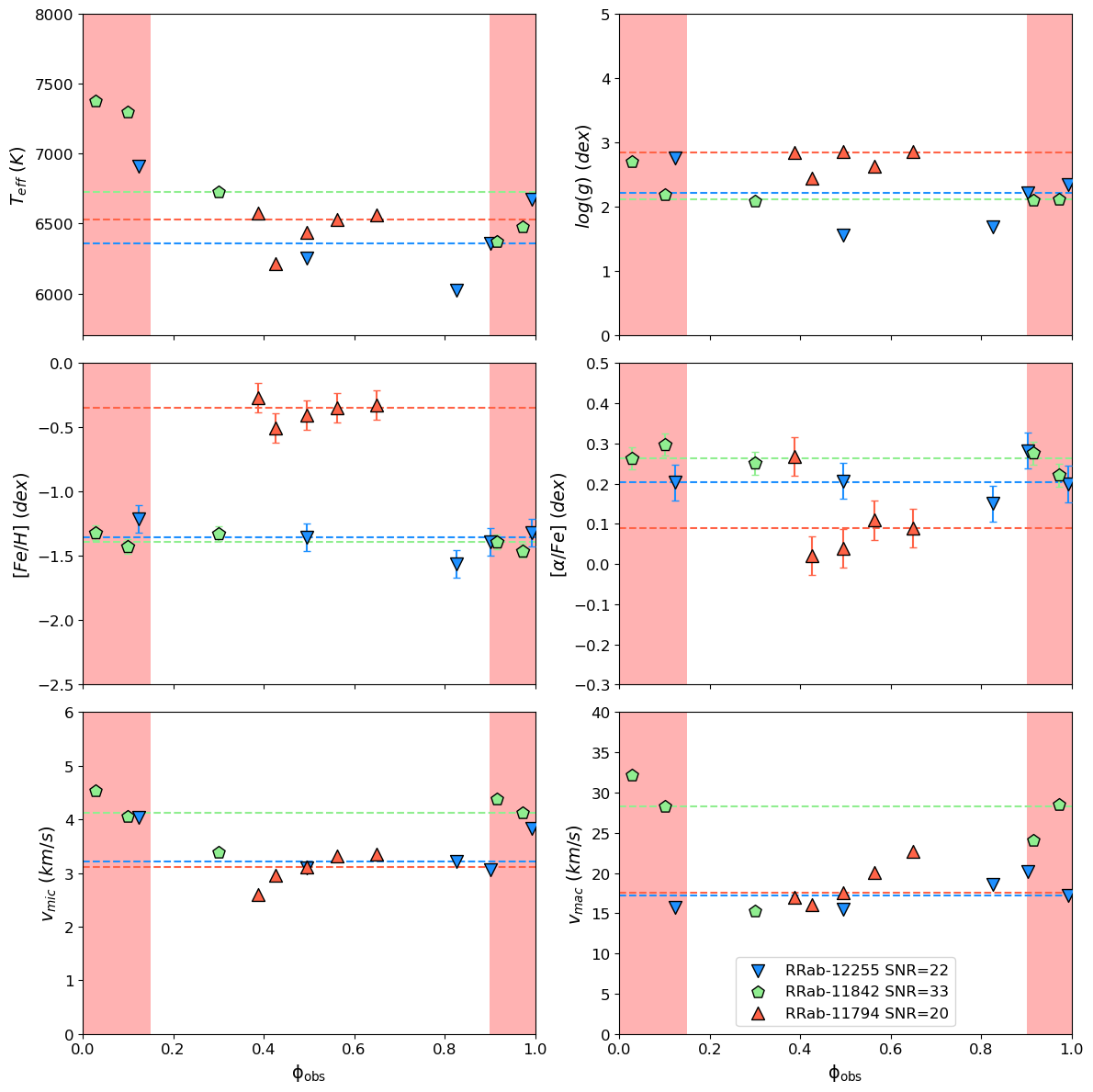}
    \caption{All the atmospheric parameter measurements per observation phase for 3 RRab stars. RRab-12255 (blue) is a metal-poor star with low SNR, RRab-11842 (green) is a metal-poor star with intermediate SNR, and RRab-11794 (red) is a metal-rich star with low SNR. Dashed lines indicate the mean value for the respective parameter for each star. Red regions indicate the main shock phase, where $\rm T_{eff}$ changes drastically.}
    \label{fig:6panel}
\end{figure*} 

The study by \cite{Prudil2025} also demonstrated that in the Galactic bulge, the Period-Luminosity relation (PLZ) exhibits a significantly stronger correlation with extinction in regions that are more heavily obscured. Fortunately, Baade's Window is characterized by low extinction levels, which allowed us to determine the distances using both the \(I\) and \(K_s\) bands, yielding very similar results. Based on the average distance distribution of our targets, we estimate the distance to the Galactic center to be \(d_{GC} = 8.17 \pm 0.11\) kpc. This finding aligns well with the accepted measurements of the Galactic center distance \citep{GRAVITY2021, Leung2023}. It is essential to note that our sample size is insufficient to constrain the Galactic center distance in a statistically robust manner. Nonetheless, our results indicate consistency with current values. The distance distributions for RRab and RRc stars are in Fig. 3 and the individual distance estimations are in Table \ref{tab:table1}.
  
\subsection{Proper Motions}
\label{subsec:PMs}

Proper motions (PMs) were obtained from the \textit{Gaia} Data Release 3 (DR3) survey \citep{GAIADR3}. The \textit{Gaia} spacecraft features a mirror measuring 1.45 m by 0.45 m and has been observing the Galaxy in the optical regime since 2014, with its mission concluding in January 2025. Previous studies have demonstrated that the precision of \textit{Gaia}'s PM measurements is well-established, with significant enhancements noted in the latest data releases. Additionally, the bulge region has been observed by \textit{Gaia}, yielding promising results for areas such as Baade's Window, which are less affected by extinction.

We initially utilized the \cite{Clementini2023} catalog of RRL variables to identify our stars through a cross-match with \texttt{Topcat} \citep{topcat}. In the cross-match we also used the magnitude as a reference and then we checked if it is the correct star comparing the periods. Our search yielded 57 RRab and 15 RRc stars from the total. This indicates that some of the RRL confirmed by OGLE-IV were not validated as RRL in \textit{Gaia} DR3. This finding is notable but not uncommon in regions such as the bulge, where crowding can significantly affect the accuracy of confirmations. For the remaining variables, we conducted another cross-match directly with the Gaia DR3 catalog, again using the magnitude as reference, which enabled us to identify the remaining stars. However, for two of the RRab stars, the catalog lacked proper motion (PM) measurements. The PMs we obtained for our sample are consistent with bulge stellar populations and are listed in Table \ref{tab:table1}.

\subsection{Systemic radial velocities}
\label{subsec:RVs}

Since RRL stars are pulsating variables, the observed RV must be corrected by the pulsation at the moment of the observation in order to obtain the real line of sight velocity. Systemic, or barycentric radial velocities ($V_\gamma$) for each star can be derived from the available GIRAFFE spectra. The estimation and correction of the radial velocity for individual spectra at various phases for a given star were conducted using an in-house code. This code cross-correlates the stellar spectrum with a small grid of synthetic templates, selecting the one with the smallest $\chi^2$ value after applying a preliminary radial velocity correction. This chosen template is then used to obtain the final estimate of the radial velocity.

To calculate $V_\gamma$, we adopted the same methodology as outlined in \cite{Prudil2024b}. They derived a radial velocity (RV) curve, and consequently, the observed RV per phase for both RRab and RRc stars using spectroscopic samples from APOGEE and \textit{Gaia}. These RV curves are suitable for our research, given that our observations are in the optical range, and they enhance the accuracy of the $V_\gamma$ estimation by 50\%. With these RV curves, we achieved consistent results for both RRab and RRc types. The calculation of $V_\gamma$ involves adjusting the RV value for the observation phase by the amplitude of the light curve. Thus, the five observation phases per star are utilized to select the observed RV and, in turn, estimate the final $V_\gamma$ at $\phi=0.37$, which indicates the moment when the star is in its mean brightness \citep{Kunder2020,Prudil2025}. The final values are listed in Table \ref{tab:table1}.
\begin{figure*}
	\includegraphics[width=\linewidth]{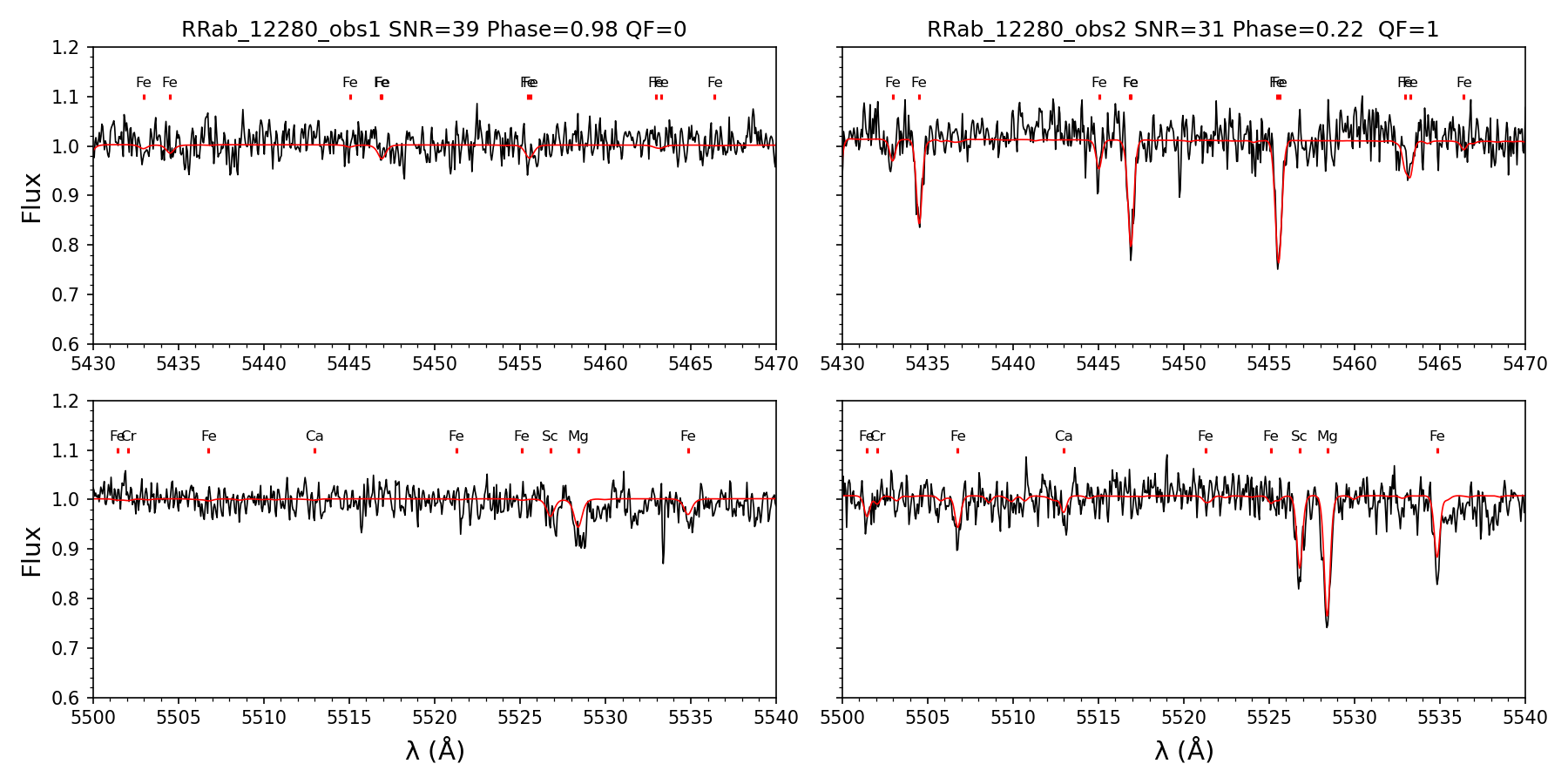}
    \caption{Two observed spectra of the star RRab-12280 and the best fit adjusted by FERRE. The normalized and corrected observed spectrum is shown in black, while the best-fit selected by FERRE is in red. Red ticks with labels indicate a set of strong lines available in this region. The spectrum on the left is the first observation of the star with SNR = 39 at $\phi_{obs}$ = 0.98; this spectrum does not show prominent lines produced by the shock in the RRL star. At the right, the second observation for the same star with SNR = 31 and $\phi_{obs} = 0.22$ is shown. In this phase, absorption lines become visible.}
    \label{fig:QF}
\end{figure*}

\section{Iron and $\alpha$-element abundances}
\label{sec:metalpha}
\subsection{Full spectral fitting}
\label{subsec:fullspectral}

We estimate atmospheric parameters through full-spectrum fitting of the observed spectra against a grid of synthetic spectra. We used the analysis code \texttt{FERRE} \citep{FERRE}. \texttt{FERRE} matches models to data, identifying the model parameters that best reproduce observations. We used the \texttt{Turbospectrum} radiative transfer code \citep{TURBOSPECTRUM}, the MARCS stellar atmospheric models \citep{MARCS}, and the \textit{Gaia} ESO survey line list \citep{GAIAESOLINE} to create the synthetic library. Furthermore, we calculate atmospheric models covering appropriate ranges for the atmospheric parameters: effective temperature ($\rm T_{eff}$), surface gravity ($\log{g}$), metallicity ($\rm [M/H]$), $\alpha$-element abundance ($\rm [\alpha$/M]), microturbulence velocity ($\rm v_{mic}$). At the moment of compiling the grids for \texttt{FERRE}, one extra dimension was added for the macroturbulence velocity ($\rm v_{mac}$). 

A set of three main grids of models covering the regions around the HB of the HRD was produced. The models are in the same spectral region as the HR10 mode of GIRAFFE and have the exact resolution. Depending on the region covered, these grids were referred to as \textit{main}, \textit{warm}, and \textit{hot}.

The \textit{main} grid comprises the region of the H-R Diagram around $\rm T_{eff}=4000-5500 \ K$ and $\log{g}=0-5.5$. It is the location where we can generally find Red Giant Branch (RGB) and Red Clump (RC) stars. The \textit{warm} grid is extended between $\rm T_{eff}=5000-7500 \ K$ and
$\log{g}=1-5.5$. This region is where HB stars are generally found, and it is the location where the majority of RRLs reside, as they orbit around the Instability Strip, which is situated in this region. Finally, the \textit{hot} grid is arranging from $\rm T_{eff}=5500-8000 \ K$ and $\log{g}=2-5.5$. This region is made for hotter stars, such as RRc, and would be useful for those. 

The analysis consisted of compiling the individual spectra, resampling the fluxes to be cast to the wavelength sampling of the grids, and running them against the three grids. \texttt{FERRE} interpolates in the grid in order to minimize the $\chi^2$ between the observed and synthetic fluxes, normalizing both with a fourth-grade polynomial. The parameters of the best-fit model are reported in one of the files, and we assume that they are the best estimation of the analyzed star. \texttt{FERRE} also creates another file that saves the best-fit model, and we selected the option to save the re-normalized version of the spectra. The best $\chi^{2}$ result between the different grids was chosen as the best model for that spectrum. Fig.~\ref{fig:spec2} shows an example of one spectrum and the corresponding best-fit model for an RRab star. The \textit{warm} grid was generally the best for RRab and RRc stars, while the \textit{hot} grid was selected for some RRc stars.

In order to estimate realistic errors for $\rm [Fe/H]$ and $\rm [\alpha$/Fe], we took advantage of the fact that we have five spectra, at different pulsation phases, for each RRL star. Therefore, for each of those spectra, we performed 1000 flux Markov-Chain Monte Carlo (MCMC) flux resamples, assuming a Poisson of the flux error. Then, we re-derived the atmospheric parameters, including metallicity. We obtained variations with sigma of 0.08 dex for $\rm [Fe/H]$ and 0.04 dex for $\rm [\alpha$/Fe]. Fig.~\ref{fig:cornerplot} shows a corner plot with the results of the MCMC for the fundamental parameters of one RRL star as an example. In addition, Appendix ~\ref{sec:cornerplot} contains a description of the observed correlations.

In order to analyze the consistency of the measured atmospheric parameters, we created diagnostic plots to test the results for different parameters versus the observation phase for each spectrum, considering that we have five observations per star. The observation phases were obtained using the period and the time of the maximum of the light curve from the OGLE-IV catalog, combined with the MJD of the observation in the header of each spectrum. Fig.~\ref{fig:6panel} shows the result for three RRab stars and their six atmospheric parameters. Fig.~\ref{fig:6panelRRc} shows the same for two RRc stars. The results are consistent with those of previous studies \citep{Pancino2015}. For instance, we observed that $\rm T_{eff}$ changes with the observation phase for each star, even for different SNR, while $\log{g}$, $\rm [Fe/H]$, and 
($\rm [\alpha$/Fe] do not change significantly with phase. This result is crucial in determining the mean value of each fundamental parameter. Consequently, $\rm v_{mic}$ and $\rm v_{mac}$ are naturally changing with phase. Since these two parameters indicate physical processes occurring in the star's atmosphere, it is normal for them to change with the star's pulsation. In summary, we conclude that the atmospheric parameters are well-constrained using the full spectral fitting method.

Fig.~\ref{fig:6panel} also shows that stars with different SNR do not produce significant changes in the obtained fundamental parameters. Moreover, we do not observe a difference in the dispersion. We will call metal-poor (m-poor) RRab those with $\rm [Fe/H]<-1$ dex and metal-rich (m-rich) those with $\rm [Fe/H]\geq-1$ dex following the separation of previous studies \citep{Du2020,Prudil2025}. From the same figure we observe that m-rich RRL have consistently higher $\log{g}$, and lower $\alpha$-element abundances than the m-poor ones.

\subsection{Quality selection}
\label{subsec:QF}

In order to obtain the mean values of the fundamental parameters for each star, we decided to first look at the fitting adjustments. As previous studies have stated \citep{Pancino2015} (hereafter, P15), there are certain regions during the observation phase where the absorption lines undergo drastic changes due to the pulsation of the stars. Sometimes, the change is significant, and the absorption lines are almost lost. By the time of the observations, these stars were observed at random observation phases, so this factor could be significant. 

We define a quality factor (QF) to select the reliable synthetic fits to determine the fundamental parameter mean values. We focused on two regions of the spectra to determine this quality factor. The former, between 5430 and 5470\AA  \ where we can see five iron I (FeI) lines. From these lines, three are particularly strong and valuable in determining whether the fitting model is good. We examine this region to ensure that our estimation of the metallicity is reliable. The latter region is from 5500 to 5540 \AA . This region contains FeI lines, too, but more importantly, it contains magnesium I (MgI) and calcium I (CaI) lines. We have one MgI line, but it is one of the strongest absorption lines in this spectral range. Additionally, there is a small CaI line, which is the most abundant $\alpha$-element in this spectral region. Thus, the second region is useful to test our $\rm [\alpha$/Fe] estimations. 

Fig.~\ref{fig:QF} shows an example of both scenarios occurring in the same star. In the right panel, we display a spectrum with QF=1, meaning that the metallicity and $\rm [\alpha$/Fe] ratio are both reliable, and in the left panel, we display the other scenario, where the absorption lines are almost lost, and no information can be obtained from the spectrum, so QF=0. The observation phase is then crucial in defining these regions, but there are also some particular cases to consider.

Using the QF, we found a star with ID OGLE-RRLYR-BLG-12573, which shows no variability. This is because the five available spectra show the same RV at distinct observation phases and no change in the absorption line shape. It is also a considerably more m-rich star. After the FERRE run, the star shows parameters related to an RC star, so it was discarded for further analysis. We believe that, as the star was discovered by OGLE using a small telescope, when we observed it with the VLT, a considerably larger telescope, more stars were resolved, and the fiber was accidentally placed on a neighboring star in this crowded region.

We found two RRc stars that show similar patterns to the previous one, with names OGLE-BLG-RRLYR-11456 and OGLE-BLG-RRLYR-12432. These stars exhibit negligible changes in RV versus observation phase, but show some differences in the spectra. We observed both in Aladin \citep{aladin1} using a VVV map and found that both stars are blended. This blending can explain the RV estimations, so these two are also discarded for further analysis.

After the quality selection, we have 60 RRab and 18 RRc stars with robust values for their atmospheric parameters. Considering the stability of iron and $\alpha$-element abundances, we calculate the mean values for both parameters for each star, which are very useful for studying the evolution of stellar populations. 

\subsection{Mean values for iron and $\alpha$-element abundances}
\label{subsec:fundparams}

We obtain median values for the RRab and RRc stars. The RRab stars have a median metallicity of $\rm [Fe/H]_{median}=-1.34 \pm 0.04$ dex, where the error is the standard deviation of the distribution divided by the square of the number of stars. For RRab $\alpha$-element abundance we obtained a median of $\rm [\alpha/Fe]_{median}=0.25 \pm 0.01$ dex. In the case of the RRc the values are $\rm [Fe/H]_{median}=-1.44 \pm 0.08$ dex and $\rm [\alpha/Fe]_{median}=0.24 \pm 0.03$ dex. The metallicities for RRab and RRc stars and their difference, where RRc stars are systematically more m-poor than RRab stars, agree well with previous spectroscopic estimations, where this difference is explained by differences in the evolutionary path \citep{Fabrizio2019,Crestani2021,Crestani2021b}.

\begin{figure}
	\includegraphics[width=\linewidth]{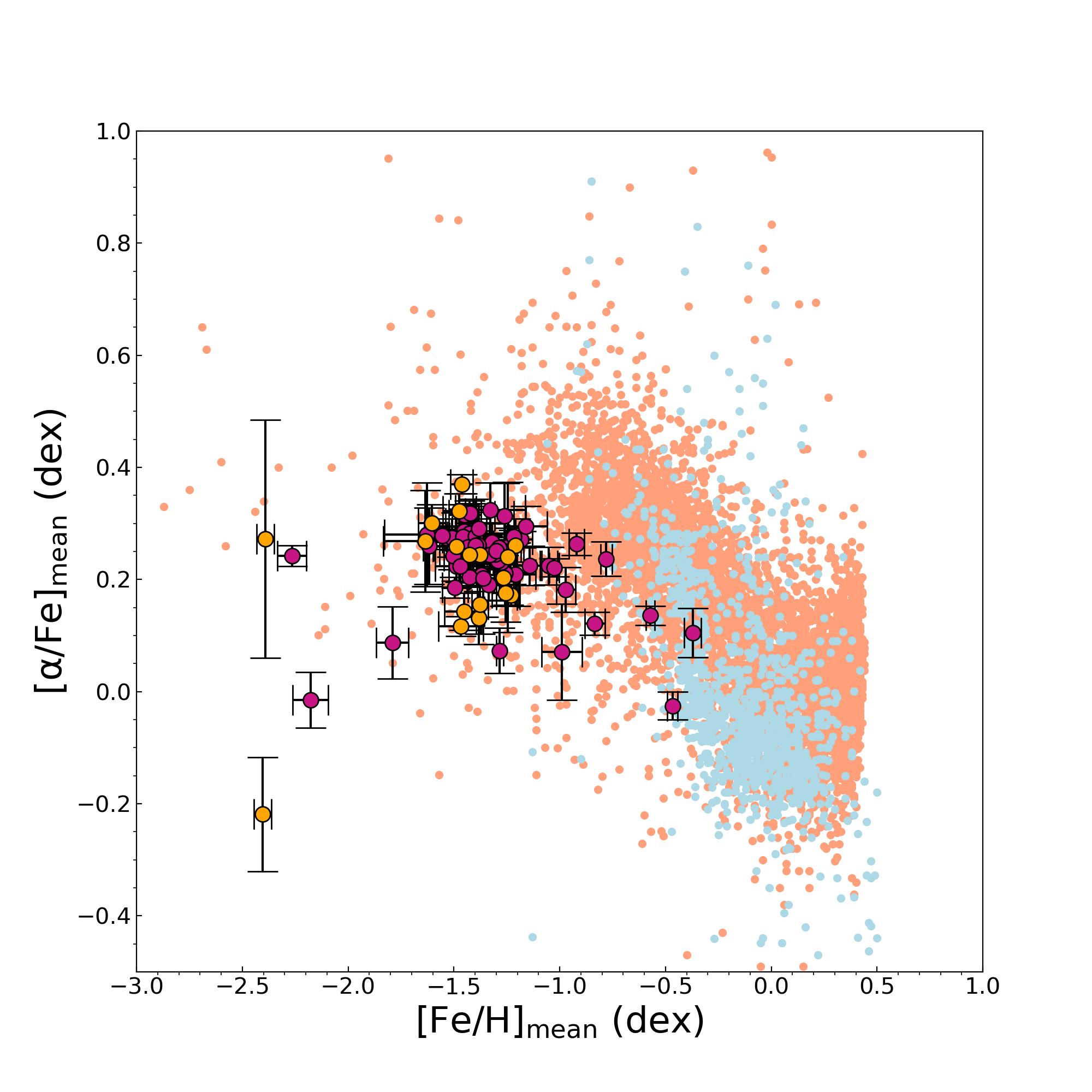}
    \caption{Abundance of $\alpha$-elements as a function of the metallicity. The red-violet circles are the RRab stars, and the orange circles are the RRc ones. Light red and light blue dots are bulge and disk giants, respectively, from the GES DR4 survey, for comparison.}
    \label{fig:Fehvsalpha}
\end{figure}

\begin{figure*}
	\includegraphics[width=\linewidth]{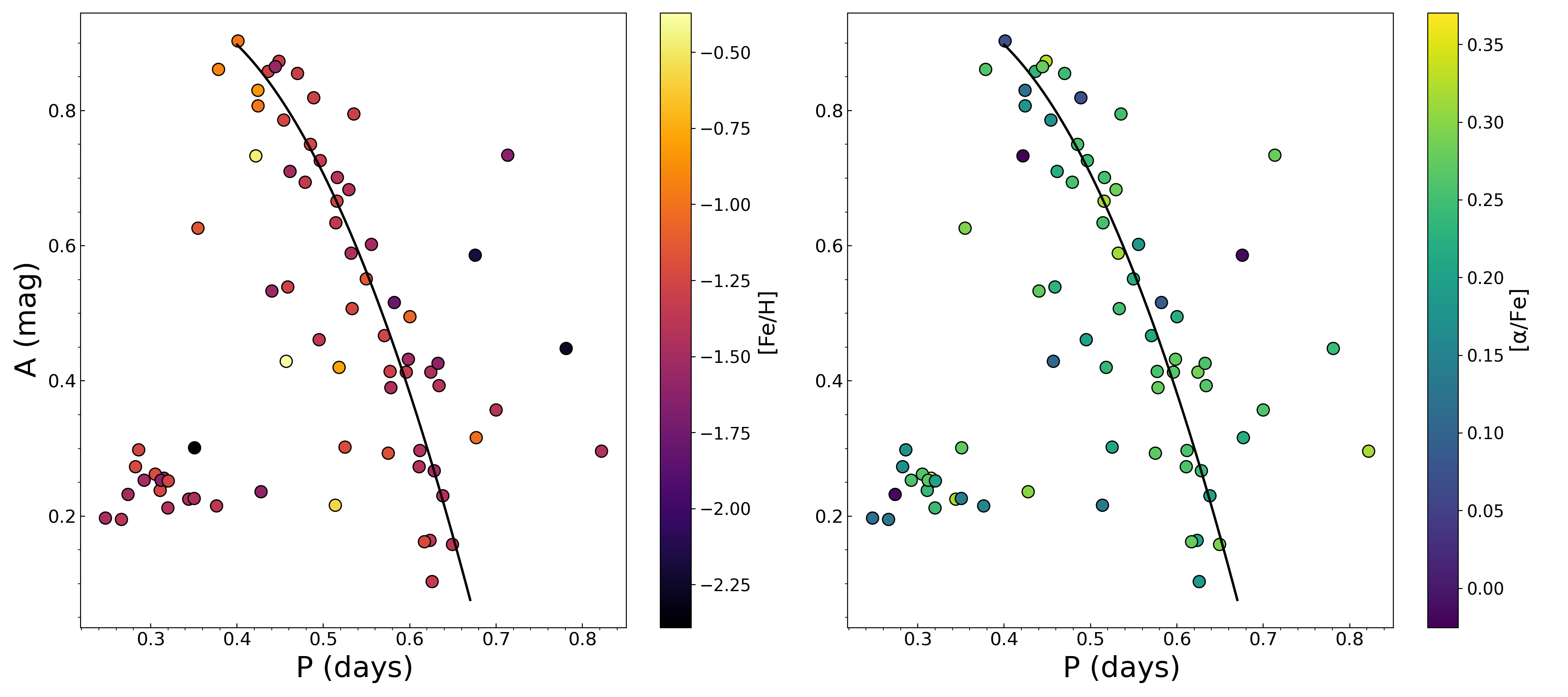}
    \caption{Bailey diagram of the RRL sample of this study. Left panel: The Bailey diagram colored by metallicity. The black line is the region of the Oosterhoff group I obtained using the stars in that region. Right panel: The Bailey diagram colored by $\alpha$-element abundance.}
    \label{fig:baileyfeh}
\end{figure*}

\begin{figure*}
	\includegraphics[width=\linewidth]{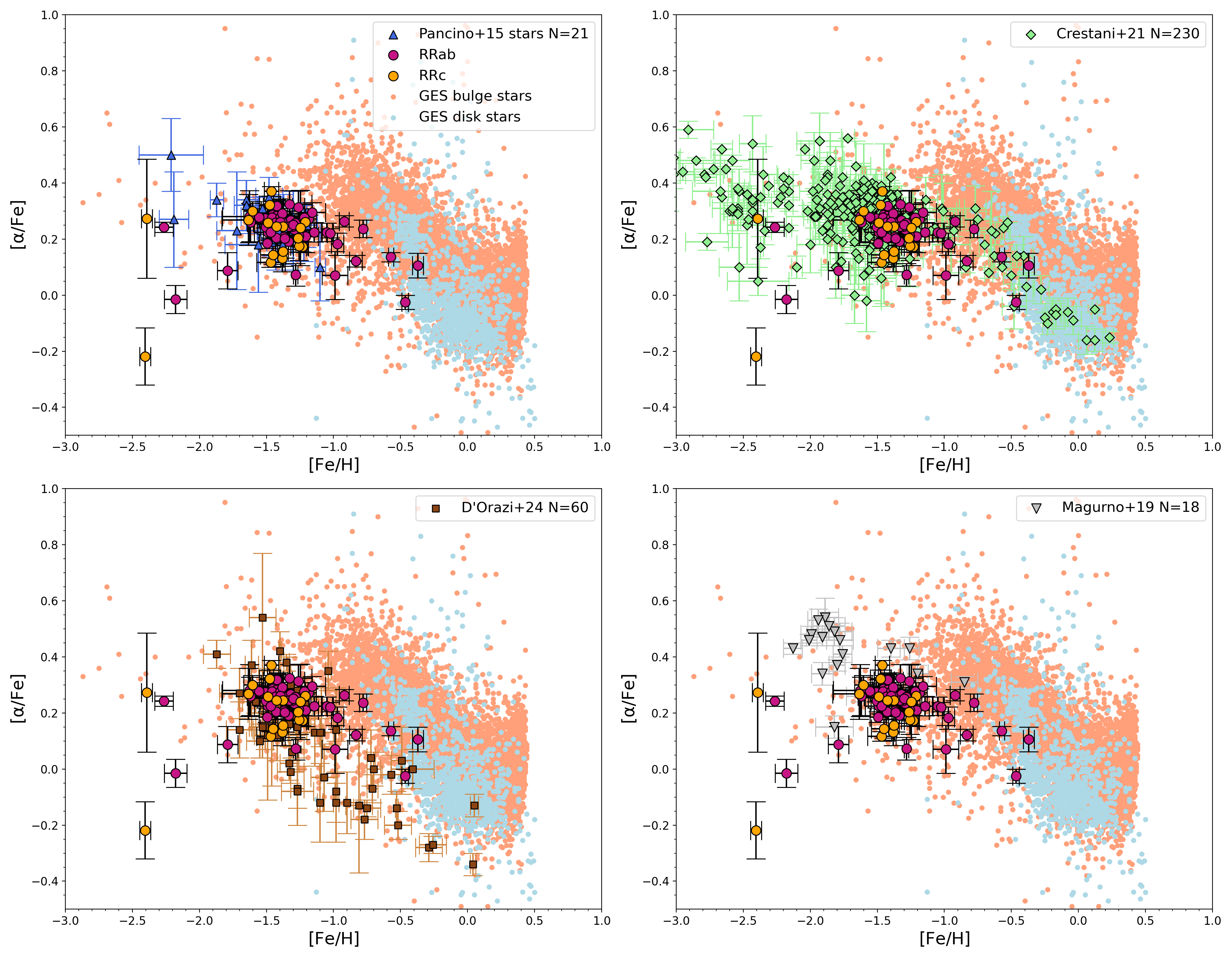}
    \caption{$\alpha$-element over iron ratio as a function of metallicity, compared with previous studies. In all the panels, the red violet circles represent the RRab stars, and the orange circles represent the RRc ones. Light red and light blue dots are bulge and disk giants, respectively, from the GES DR4 survey for comparison. \textit{Top left panel:} blue triangles are RRL stars from \cite{Pancino2015}. \textit{Top right panel:} Light green diamonds are RRL stars from \cite{Crestani2021b}. \textit{Bottom left panel:} Brown squares are RRL stars from \cite{DOrazi2024}. \textit{Bottom right panel:} Grey triangles are RRL stars from \cite{Magurno2019}.}
    \label{fig:Fehvsalpha2}
\end{figure*}

We found a difference with the only previous work focused on high-resolution spectroscopic metallicities of RRL stars in the bulge \citep{Walker1991}, where they found a mean metallicity of $\rm [Fe/H]=-1.05 \pm 0.16$ dex for 41 RRab stars. However, our result is in agreement with that of \cite{Savino2020}, which used the Calcium triplet method to obtain spectroscopic metallicities and found a mean of -1.24 dex. We note that at least part of the offset can be explained by the different metallicity scale. Indeed, \cite{Crestani2021}, calculated a difference of 0.08 dex between the [Fe/H] they derive based on the scale of \cite{For2011}, \cite{Chadid2017}, \cite{Sneden2017} (hereafter FCS) and those of \cite{Carretta2009}. Assuming the same applies to our measurements, the latter would be 0.08 dex higher on the scale by \cite{Carretta2009}. Correcting by this factor, the mean metallicity for our RRL sample would be $-$1.26 dex, in agreement with \cite{Savino2020}.   

Figure~\ref{fig:Fehvsalpha} shows the [Fe/H] vs $\rm [\alpha/Fe]$ plane, which is crucial to analyze the evolutionary path of the RRLs. The errors shown are the standard error of the mean of the obtained values for each observed phase of a single star, which already include the individual error on the parameter. The time delay model explains that supernovae II enrich the interstellar medium (ISM) earlier, since massive stars collapse rapidly with time. They provide $\alpha$-elements and iron with a fixed proportion to the ISM. After some time, supernovae Ia enrich the ISM principally with iron peak elements, producing a decrease in the $\rm [\alpha/Fe]$ ratio, so new generations of stars have less $\rm [\alpha/Fe]$ and higher levels of [Fe/H]. The light red dots in Fig.~\ref{fig:Fehvsalpha} are abundances of bulge stars from the Gaia ESO Survey Data Release 4 (GES DR4), while light blue points are disk stars from the same survey. Our RRLs are primarily m-poor, with a peak near -1.35 dex, but there are some RRab that could be considered m-rich, with metallicities greater than -1 dex. Some of these stars also show lower abundances of $\rm [\alpha/Fe]$. There are no m-rich stars in the RRc group.

The location of most RRL stars is at $\alpha$-element abundances around 0.25 dex. This value is in the range of the bulge, although it is slightly lower compared with other bulge m-poor stars. There are two possible explanations. Firstly, our $\alpha$-element abundances are based on Ca and Mg lines, where previous studies show that calcium abudance is generally lower than other $\alpha$-elements in RRL stars \citep{Pancino2015,DOrazi2024}. This hypothesis could be tested by means of high resolution spectra covering a wider wavelength range, therefore allowing the direct measurement of different elements. The other possibility is that bulge RRLs have a different star-formation history, meaning they formed under different conditions in the interstellar medium at an early stage of the Galaxy. 

We found that our m-rich RRL stars exhibit $\rm [\alpha/Fe]$ ratios that vary with metallicity, suggesting that this group may have mixed origins. As a test, Fig. ~\ref{fig:baileyfeh} shows the period and amplitude of our sample colored by metallicity in the left and by $\alpha$-element abundance at the right. From these figures, the majority of the RRab stars are located around the black line, calculated using a spline fit to the stars in that region, and this is the typical tendency of the Oosterhoff I (OoI) group for RRL stars. The stars on the right side are likely RRab stars associated with the Oosterhoff II (OoII) group. A surprising point is that several of the m-rich RRab stars with lower levels of [$\alpha$/Fe] are located at the left side of the OoI group. \cite{Prudil2025} suggests that a possible cause for the different metallicities of RRL stars is a difference in age of about 2 Gyr, whereby the oldest RRL could have ages around 12 Gyr, while the "younger" ones have ages around 10 Gyr. Our results suggest that this working hypothesis may be true; however, further data on m-rich RRL stars in the bulge are needed to draw a stronger conclusion. Further discussion about these m-rich RRL stars will be given in Section~\ref{sec:chemodynamics} below, when analyzing orbits.

\begin{table*}[]
\centering
\scriptsize
\caption{Observational parameters of the RRLs in our sample and the ones derived in this work.}
\begin{tabular}{llllllllllllll}
\begin{tabular}{lllllllllllll}
\hline
OGLE ID & $\alpha$   & $\delta$   & Type & $P$        & $\langle I \rangle$ & $A_I$ & $V_\gamma$         & $\rm [Fe/H]$ & $\rm [\alpha/Fe]$ & $D$   & $\mu_\alpha$          & $\mu_\delta$          \\
        & (deg)      & (deg)      &      & (days)     & (mag)               & (mag) & $\rm (km\ s^{-1})$ & (dex)              & (dex)                     & (kpc) & ($\rm mas \ yr^{-1}$) & ($\rm mas \ yr^{-1}$) \\ \hline  
11978   & 270.821249 & -29.956638 & RRab & 0.64948819 & 15.853              & 0.158 & 2.2                & -1.38              & 0.29                      & 9.4   & -6.18                 & -3.24                 \\
11763   & 270.706708 & -29.739055 & RRab & 0.42178368 & 16.183              & 0.733 & 273.5              & -0.46              & -0.02                     & 9.5   & -3.65                 & -7.95                 \\
12894   & 271.324874 & -28.782888 & RRab & 0.67582229 & 15.569              & 0.586 & 64.7               & -2.17              & -0.01                     & 8.1   & -8.55                 & -6.11                 \\
12925   & 271.342916 & -28.725333 & RRab & 0.53328844 & 15.735              & 0.507 & 19.9               & -1.23              & 0.25                      & 7.6   & -5.15                 & -5.02                 \\
12753   & 271.261958 & -28.772861 & RRab & 0.51380894 & 15.705              & 0.216 & 79.5               & -0.57              & 0.13                      & 7.2   & -1.56                 & -0.40                 \\
12768   & 271.270124 & -28.647499 & RRab & 0.57512043 & 15.672              & 0.293 & -118.2             & -1.18              & 0.27                      & 7.8   & -2.25                 & -0.19                 \\
12502   & 271.122041 & -28.849638 & RRab & 0.62836354 & 14.766              & 0.267 & 495.2              & -1.50              & 0.24                      & 5.8   & 1.24                  & -9.52                 \\
11794   & 270.723458 & -29.646361 & RRab & 0.45696349 & 15.577              & 0.429 & 142.5              & -0.37              & 0.10                      & 6.4   & -5.26                 & -4.08                 \\
11553   & 270.611333 & -29.678166 & RRab & 0.63828087 & 15.465              & 0.230 & 133.1              & -1.42              & 0.20                      & 7.1   & -5.26                 & -3.97                 \\ \hline
\end{tabular}
\end{tabular}
\begin{tablenotes}
\item All IDs start with the OGLE-BLG-RRLYR identifier. The coordinates, type, periods, mean magnitudes, amplitudes, velocities, metallicities, $\alpha$-element abundances, distances, and proper motions are presented. The complete version, which includes the \textit{Gaia} IDs and the errors, is only available in electronic form at the CDS via anonymous ftp to cdsarc.u-strasbg.fr (130.79.128.5) or via http://cdsweb.u-strasbg.fr/cgi-bin/qcat?J/A+A/.
\end{tablenotes}
\label{tab:table1}

\end{table*}

\section{Comparison with previous studies}
\label{sec:comparingfe}
\begin{figure*}
	\includegraphics[width=\linewidth]{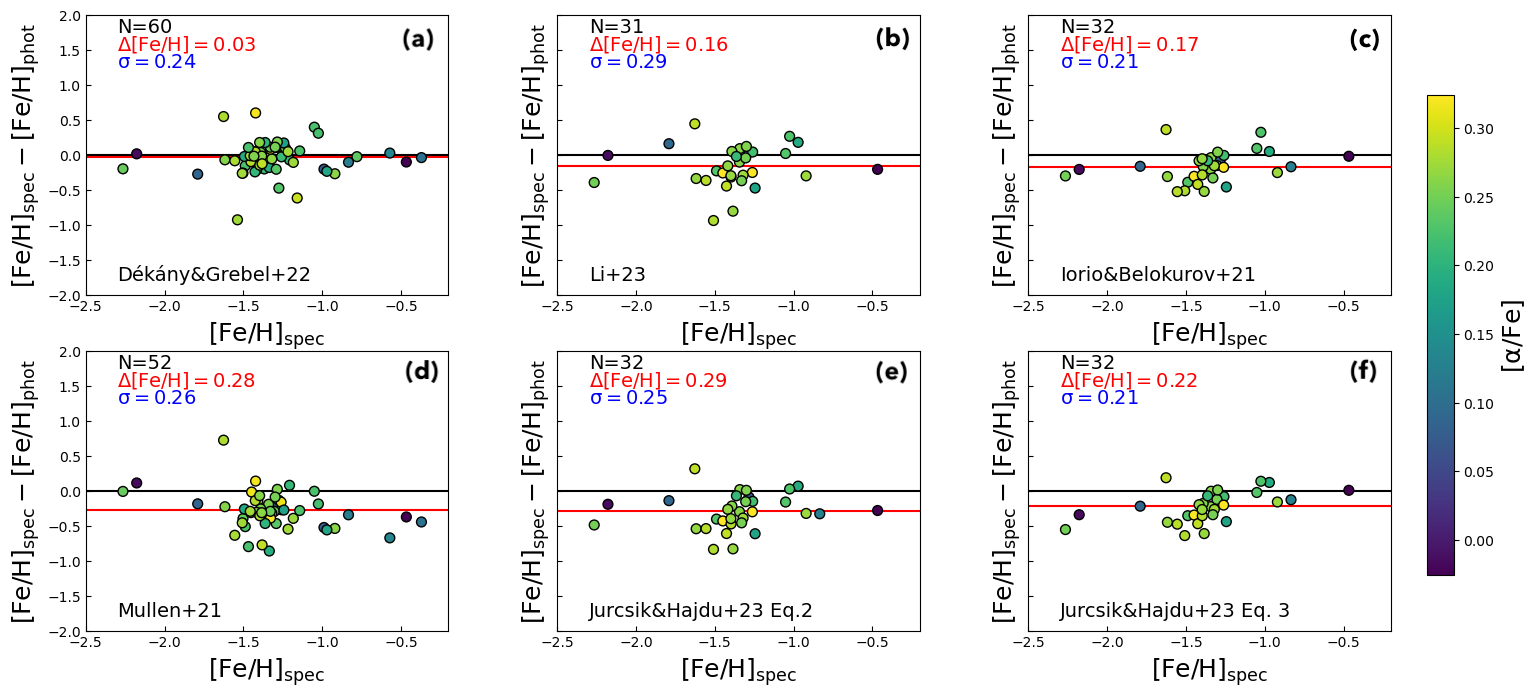}
    \caption{Spectroscopic metallicity compared with the photometric metallicity of recent studies for the RRab stars. The circles are color-coded by $\alpha$-element abundance. From top left to bottom right, the studies for the comparison are \cite{Dekany2022}, \cite{Li2023}, \cite{Iorio2021}, \cite{Mullen2021}, and \cite{Jurcsik2023}  Eq.~2 and Eq.~3. The black line indicates a zero difference, and the red line the metallicity offset.}
    \label{fig:specvsphotmet}
\end{figure*}

\begin{figure*}
	\includegraphics[width=\linewidth]{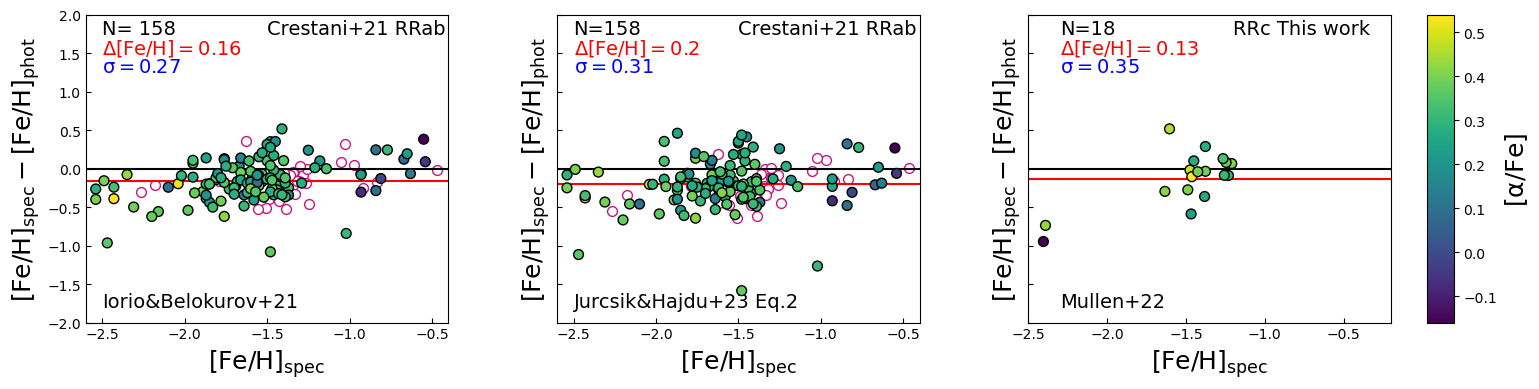}
    \caption{Spectroscopic metallicity compared with the photometric metallicity of recent studies.\textit{ Left panel}: same as Fig.~\ref{fig:specvsphotmet} but for an RRab set of \cite{Crestani2021} and compared with \cite{Iorio2021}. The circles are color-coded by $\alpha$-element abundance based in \cite{Crestani2021b}. The unfilled red-violet circles are our RRab stars. \textit{ Middle panel}: The RRab set of \cite{Crestani2021} compared with \cite{Jurcsik2023} Eq.~2. \textit{ Right panel}: The metallicity difference of our RRc stars compared with \cite{Mullen2022}.}
    \label{fig:specvsphotmetcrest}
\end{figure*}

\subsection{Comparison with previous spectroscopic abundances of RRL stars}
\label{subec:comparingphotfe}

In Fig.~\ref{fig:Fehvsalpha2} we compare our results with previous spectroscopic works that derived [Fe/H] and either the global [$\alpha$/Fe] or a single element such as Mg or Ca. They all refer to RRLs in different galactic components than the bulge.
The blue triangles in the top left panel came from P15, based on solar vicinity RRL stars, observed with SARG@TNG and UVES@VLT; they obtained iron, Mg and Ca abundances by the equivalent width method. Only the Calcium abundance is used in this figure since it is available for more stars. Our RRLs span the same region as those in P15, with the exception of the m-rich RRL variables that are not present in P15. This could be a real, evolutionary difference, but also simply low number statistics. Larger samples are needed to reach a solid conclusion.

The light green diamonds on the top right panel come from \cite{Crestani2021b} (hereafter, C21) and are RRL stars (both RRab and RRc) in the solar vicinity. Iron abundances were obtained by the $\Delta S$ method, on high-resolution spectra from the echelle spectrograph at the Du Pont Telescope (Las Campanas Observatory). The $\alpha$-element abundances, instead, were obtained by the equivalent width method. The bulk of our RRLs share the same abundance ratios as those in C21, however, our sample does not include RRLs as m-poor and $\alpha$-rich as those of C21 suggesting a real lack of those stars in the Galactic bulge, compared to the local disk.

The brown squares on the bottom left panel are solar vicinity RRL stars from \cite{DOrazi2024} (hereafter, D24). They obtained abundances for several alpha elements by the full spectral fitting method on high-resolution spectra. We show only Calcium abundances here, because they are the most robust measurements in that study. Since our $\alpha$-element abundances were based mainly on Ca and Mg lines, this comparison is appropriate. The bulk of our RRLs overlap with the D24 sample. The m-rich RRLs from D24, however, have a lower [$\alpha$/Fe] ratio compared with ours. These lower values were not expected for the MW disk, which is why they propose a different origin (a primordial disk) for these stars.

Finally, the bottom right panel shows, as grey triangles, RRLs from \cite{Magurno2019} (hereafter, M19). These variables belong to the globular cluster $\omega$ Cen. M19 used high-resolution spectra to obtain abundances using the equivalent width method for about 80 RRL stars, but they provide $\alpha$-element values for only 18 RRL stars. As expected, our abundances differ from those of M19 in almost the entire distribution. $\omega$ Cen RRL stars are more m-poor and $\alpha$-enhanced than bulge RRL variables. A plausible reason for this difference could be that the population of $\omega$ Cen, located in the Galactic halo, has intrinsically less metal content due to a different chemical evolution. Furthermore, it is essential to note that the abundance analysis method employed in M19 differs significantly from the one used in this study; therefore, the abundance scales may have an offset.

\begin{figure*}
	\includegraphics[width=\linewidth]{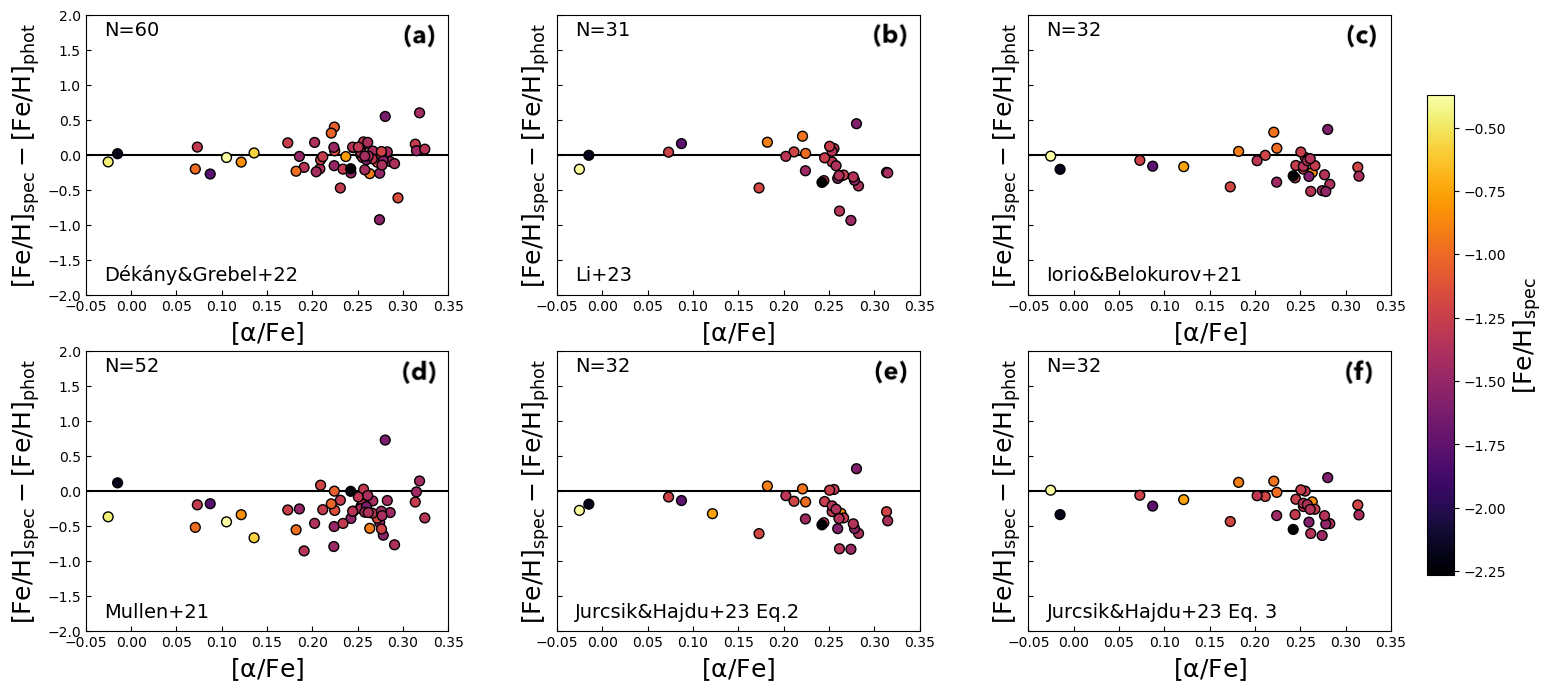}
    \caption{Difference in spectroscopic and photometric metallicities as a function of the $\alpha$-element abundances for the RRab stars. The circles are color-coded by metallicity. The comparison is using the same studies as for Fig. ~\ref{fig:specvsphotmet}. The black line indicates a zero difference.}
    \label{fig:alphavsphotmet}
\end{figure*}

In summary, the [$\alpha$/Fe] ratio for our RRL variables in the bulge are consistent with those of previous studies. The observed differences are compatible with those expected in different Galactic components.

\subsection{Comparison of spectroscopic vs photometric metallicities for RRL stars}
\label{subec:comparingphotfe}

One of the primary goals of this study is to provide a sample of RRLs in the bulge to be used as calibrators, or at least a reference, to validate the different methods proposed to derive metallicity from photometry.

To this end, we provide here a comparison between our results and several photometric methods recently proposed. Specifically,
we derived photometric metallicities for our RRLs by means of $K_s$-band light curves, as proposed by \cite{Dekany2022} (hereafter DG22), by means of the $g$-band light curves from the \textit{Gaia} DR3 survey according to the prescriptions by 
\cite{Li2023} (hereafter Li23) or by \cite{Iorio2021}, by means of V-band lightcurves as proposed by \cite{Mullen2021} (hereafter M21), and, finally, by using the periods and $\varphi_{31}$ as in \cite{Jurcsik2023} (hereafter JH23) using their Eq.~2 or their Eq.~3.

The $K_s$-band light curves for 60 of our variables were extracted from the VVV survey, corrected by the Heliocentric Julian Day (HJD), and then feed to the \textit{lcfit}\footnote{https://github.com/idekany/lcfit} code to be phased. Afterwards, we use the \textit{rrl\_feh}\footnote{https://github.com/idekany/rrl\_feh\_nn} code to obtain the photometric metallicities. The method is based on a set of light curves from VVV, used as a training set. The differences between these metallicities and our spectroscopic measurements are shown in Fig.~\ref{fig:specvsphotmet}, panel-a.

There are 31 RRLs in common between our catalog and Li23. For these stars we retrieved their photometric metallicity directly from the catalog available in CDS. Their method is based on the $g$-band light curves from Gaia DR3, together with the periods and the $\varphi_{31}$ parameters. The differences are in Fig.~\ref{fig:specvsphotmet}, panel-b.

In order to compare with the method proposed by IB21, we performed a cross-match between our sample and the \textit{Gaia} DR3 catalog of RRL stars \citep[][hereafter C23]{Clementini2023}. For the 32 common variables, we retrieved the periods and the $\varphi_{31}$ values from Gaia, and obtained the photometric metallicity using Eq. 3 from IB21. It is worth noticing that the $\varphi_{31}$ parameter provided in the Gaia catalog is in the $cos$ form, and it must be transformed into the $sin$ form in order to be feed into the equation. The differences are shown in panel-c of Fig.~\ref{fig:specvsphotmet}.

The method proposed by M21 is based on the $V$-band light curve, specifically from the ASAS-SN survey \citep{Shappee2014}. All our stars were selected from the OGLE-IV catalog; therefore, they all have very well-sampled $V$ and $I$-band light curves in the Johnson-Cousins system, the same used by the ASAS-SN survey. The $\varphi_{31}$ parameter provided by OGLE-IV, however, is based on the $I$-band light curves. It was transformed into the $V$-band using the relation provided by \cite[][their Eq.~6]{Skowron2016}, and then fed to Eq.~6 by M21. The differences between the resulting metallicities and our spectroscopic values are shown in 
panel-d of Fig.~\ref{fig:specvsphotmet}.

The method proposed by JH23 is also based on the Gaia light curves and parameters, which were retrieved from C23. We fed the periods and $\varphi_{31}$ values on both their Eq.~2, calibrated upon all the globular clusters with robust spectroscopic metallicities, and their Eq.~3, calibrated only upon the OoI type clusters, which should be more representative of our bulge RRLs. The differences are shown in Fig.~\ref{fig:specvsphotmet}, panel-e and panel-f, respectively.

The colorbar in Fig.~\ref{fig:specvsphotmet} adds the $\rm [\alpha/Fe]$ dimension. The black line is a zero level reference line, while the red line is the mean difference. As expected, different methods show different mean residuals, however, with the only exception of D22, the photometric metallicities are always higher than the spectroscopic ones. This result has been observed already by \citet{Mullen2021} and by \citet{Kunder2024}, who compared with the results from the high-resolution spectroscopic measurements by C21. Some of the methods also exhibit trends with metallicity, however those are strongly influenced by the very small number of variables in the highest and lowest metallicity regimes, therefore, with the present sample, they are not statistically robust.  

Fig.~\ref{fig:specvsphotmetcrest} shows a set of 158 RRab stars from C21 in the same form as Fig.~\ref{fig:specvsphotmet}, compared with the IB21 (left) and with JH23 (right). The red-violet circles are our RRab stars. From both panels, we observe an offset in metallicities, indicating that the C21 spectroscopic metallicities are lower than the photometric ones. For the C21 RRab set, the offset is 0.16 dex when comparing with IB21 and 0.2 dex with JH23; these values are in agreement with our results (0.17 for IB21 and 0.22 for JH23 Eq.~3, respectively). Thus, using different spectroscopic datasets, we obtain similar offsets between spectroscopic and photometric metallicities.

There is no clear explanation for the offset observed. Nonetheless, a clue is the behavior of the offset for RRc stars. There are fewer photometric relations for RRc stars; however, the  \cite{Mullen2022} relation is a robust and recent one. Fig.~\ref{fig:specvsphotmetcrest} right panel shows the metallicity difference with that study for our 18 RRc stars. The offset for RRc stars is lower compared with the one for RRab in Fig.~\ref{fig:specvsphotmet} panel-d. Furthermore, if we change to the \cite{Carretta2009} scale and consider the dispersion, this result is in agreement with this relation. This means that the offset is more prominent in the case of RRab stars. Since photometric metallicity is based on empirical values of period and $\varphi_{31}$, these could be affected by different atmospheric factors present in RRab stars and not considered, as the Non-LTE effect or the $\alpha$-element abundances.

Fig.~\ref{fig:alphavsphotmet} shows the difference between our spectroscopic metallicities and the photometric ones as a function of the [$\alpha$/Fe] ratio. The color bar shows the spectroscopic metallicity. The two calibrations based on the light curves from Gaia show an increasing offset with respect to spectroscopic values, for high values of the [$\alpha$/Fe] ratios. This might suggest that the value of the abundance ratio of the calibrators might not be appropriate for bulge RRL.
In addition, all the methods show a larger spread in the residual for higher $\alpha$-element abundances. This might be just a visual effect due to the lower statistics at low alphas, or a real trend. Larger samples are needed to clarify this point. 


\section{Orbital integration}
\label{sec:orbits}

\begin{figure*}
	\includegraphics[width=14cm]{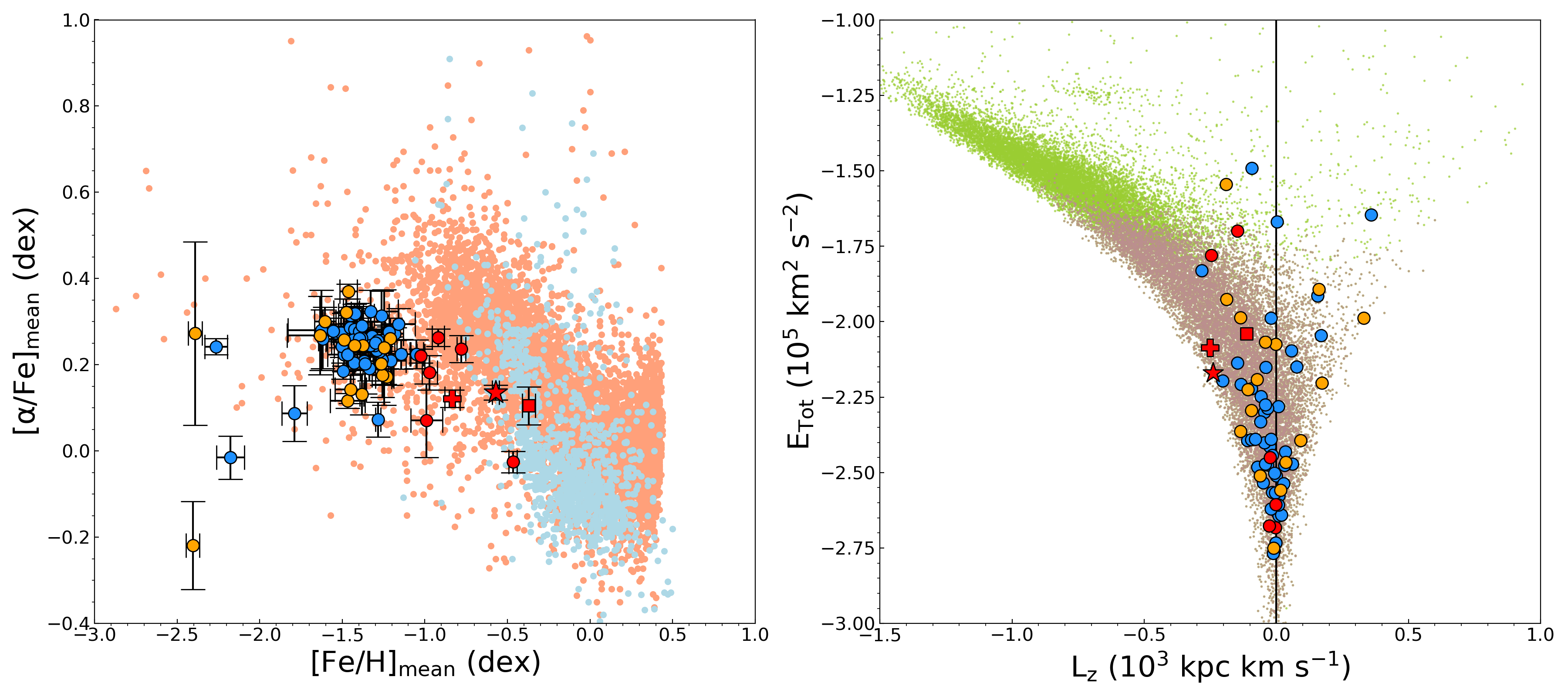}
    \centering
    \caption{Selection of the m-poor and m-rich RRab stars and their orbital distribution. \textit{Left:} The $\alpha$-abundance vs iron distribution this time separated between m-rich (red) and m-poor (blue) RRab stars, RRc stars are also included (orange). Light red and light blue dots are bulge and disk giants, respectively, from the GES DR4 survey for comparison. \textit{Right:} The $E_{tot}$ vs $L_z$ plane showing the orbital parameter results for the RRL stars with the same symbols as the left panel. Rosy-brown and light-green dots represent bulge and halo/disk giant stars, respectively, from \cite{Queiroz2023} with an orbital analysis from \cite{DeLeo2026} for a consistent comparison. The vertical black solid line separates the prograde (left) and retrograde (right) motion.}
    \label{fig:selectionandEtot}
\end{figure*}

In order to obtain the orbits of the RRL stars, we follow the same approach presented in \cite{Olivares2024}. Here, we present a brief summary of the orbital code implemented and the potentials.

We used the orbital code \textit{OrbIT} \citep{DeLeo2026}, which has the advantage of including the potentials in the inner part of the Galaxy and the possibility to change the relevance of each potential. The model of the MW gravitational potential used in our orbital integrator includes a Navarro-Frenk-White dark matter halo \citep{1996ApJ...462..563N}, two stellar and two gaseous disks with the profile by \citet{Miyamoto1975}, a rotating bar \citep{1992ApJ...397...44L}, and a spherical
bulge component modeled as a Hernquist profile \citep{1990ApJ...356..359H}. The details of the masses and all other parameters for each component of the potential can be found in \cite{DeLeo2026}.

In order to derive the orbital parameters, we converted the observed coordinates, the RVs, the PMs, and the distances into the Cartesian galactocentric frame with the Astropy modules \citep{astropy1,astropy2}. We used a distance of the Sun to the Galactic center of 8.27 kpc \citep{GRAVITY2021} and the solar velocity vector $\rm (U,V,W) = (11.1,12.24,7.25) \ km \ s^{-1}$ \citep{Schonrich2010}. The velocity of the Local Standard of Rest is assumed to be $\rm V_{LSR} = 232.8 \ km \ s^{-1}$ \citep{McMillan2017}. For each star, the orbits were evolved backward in time for 5 Gyr inside the MW potential. We obtain orbits for 57 RRab and 17 RRc stars, since 3 RRab and 1 RRc do not have PM values. The resulting orbital parameters are listed in Table \ref{tab:table2}. 

\begin{figure*}
	\includegraphics[width=\linewidth]{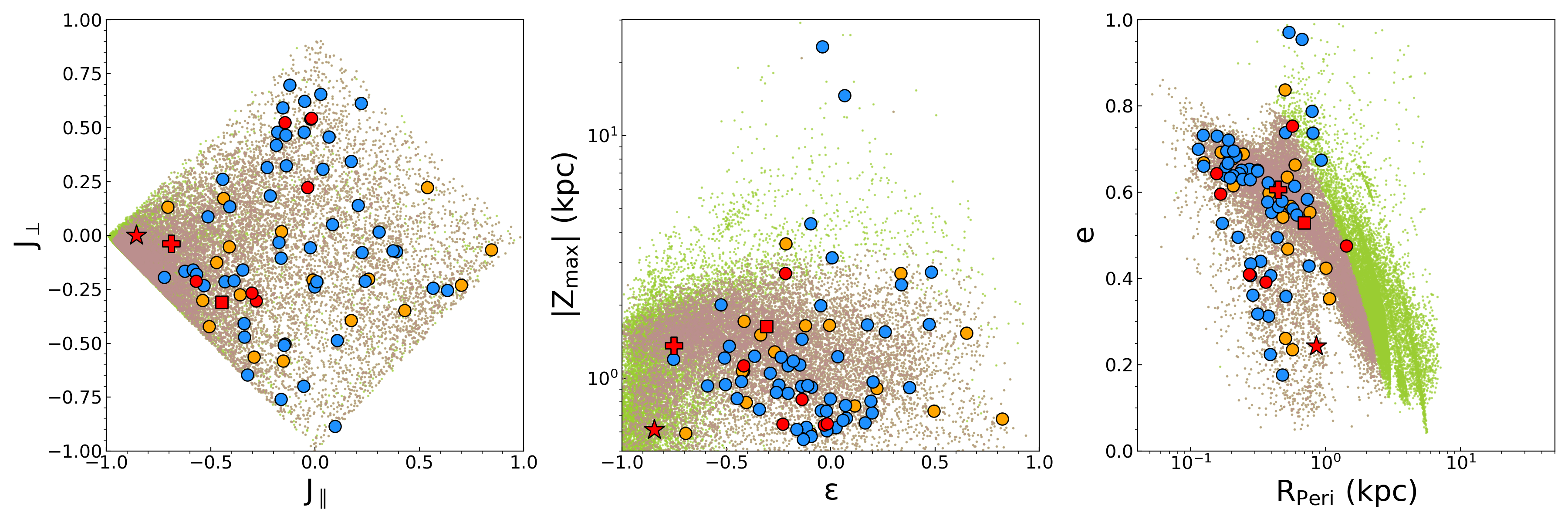}
    \caption{The location of the m-poor and m-rich RRab stars in the different orbital dimensions. For all the panels, red circles are m-rich RRab, blue circles are m-poor RRab stars, and orange circles are RRc stars. Rosy-brown and light-green dots represent bulge and halo/disk giant stars, respectively, from \cite{Queiroz2023} with an orbital analysis from \cite{DeLeo2026} for a consistent comparison. \textit{Left:} The RRL sample in the parallel versus perpendicular action space. \textit{Middle:} The RRL sample in the circularity versus maximum orbital excursion in semilogarithmic scale. \textit{Right:} The RRL sample in the pericenter versus eccentricity plane, in semilogarithmic scale.}
    \label{fig:orbitalparams}
\end{figure*}
\begin{table*}[]
\centering
\scriptsize
\caption{Orbital parameters derived for the RRLs of the sample.}
\begin{tabular}{lllllllllllll}
\hline
OGLE ID   & Classification  & {[}Fe/H{]}  & {[}$\alpha$/Fe{]}  & E$\rm _{Tot}$  & L$_z$ & $J_\parallel$ & $J_\perp$ & $\varepsilon$ & $R_A$ & $R_P$ & $\rm Z_{max}$ & $\rm e$ \\ 
& & (dex) & (dex) & ($\rm 10^5 \ km^{2} \ s^{-2}$) & ($\rm 10^3 \ kpc \ km \ s^{-1}$) & & & & (kpc) & (kpc) & (kpc)           &  \\ \hline
12255 & m-poor   &     -1.36  &   0.21 &-1.830086 &-0.283062  &-0.625 &  -0.164 & -0.526  & 2.631 &0.730 &2.0067 &0.584 \\
12280 & m-poor   &     -1.24  &   0.26 &-2.299907 &-0.0452344  &-0.324  & -0.645 & -0.205  & 1.254 &0.272 &1.1269 &0.654 \\
12285 & m-poor   &     -1.48  &   0.22 &-2.414311 &-0.0247815  &-0.163 &  -0.105  &-0.139  & 1.122 &0.216 &0.9278 &0.687 \\
12298 & m-poor   &    -1.20   &  0.21& -2.280899 &0.0072986  &0.0364  & 0.3085 & 0.0319  & 1.337 &0.398 &1.2265 &0.554 \\
12309 & m-poor   &    -1.28   &  0.07 &-2.207105 &-0.134229  &-0.514  & 0.0884  &-0.509  & 1.351 &0.371 &1.2110 &0.577 \\
12353 & m-poor   &    -1.25   &  0.21 &-2.393291 &-0.109288  &-0.585  & -0.158 & -0.591  & 1.114& 0.215 &0.9297 &0.683 \\
12376 & m-poor   &    -1.33   &  0.19 &-2.631931 &0.0029722  &0.0254 &  0.6556  &0.0245  & 0.607 &0.287 &0.6248 &0.361 \\
12403 & m-poor   &    -1.27   &  0.23 &-2.222161 &-0.0937202  & 6-0.432 &  -0.213 & -0.366  & 1.507 &0.372 &1.2337 &0.622 \\
12415 & m-poor   &    -1.34   &  0.25 &-2.643120 &0.0088065 & 0.1049  & -0.486  &0.0740  & 0.771 &0.123 &0.6849 &0.732 \\
12502 & m-poor   &    -1.50   &  0.24 &-6.37487  &-0.200613 & -0.054  & -0.698  &-0.039 &  35.93& 0.536 &23.258& 0.970 \\ \hline
\end{tabular}
\scriptsize
\begin{tablenotes}
\item All IDs start with the OGLE-BLG-RRLYR identifier. The classification, metallicity, $\alpha$-element abundance, total energy, angular momentum, parallel action, perpendicular action, circularity, apocenter radius, pericenter radius, maximum height, and eccentricity are presented. The complete version, which includes the errors, is only available in electronic form at the CDS via anonymous ftp to cdsarc.u-strasbg.fr (130.79.128.5) or via http://cdsweb.u-strasbg.fr/cgi-bin/qcat?J/A+A/.
\end{tablenotes}
\label{tab:table2}
\end{table*}

\section{Chemodynamical analysis}
\label{sec:chemodynamics}

In order to analyze the properties of the RRLs in the dynamical spaces, we divided them into two groups according to their metallicities. Specifically, we define 48 m-poor ([Fe/H]$<-$1 dex) and 9 m-rich ([Fe/H]$>-$1 dex) stars. All the c-type RRLs fall in the m-poor group.

The left panel of Fig.~\ref{fig:selectionandEtot} shows the $\rm [\alpha/Fe]$ vs [Fe/H] plane for the RRab stars, separated into m-poor (blue) and m-rich (red), RRc stars are also included in orange. We used different markers (square, cross, star) to identify three m-rich RRLs with $\rm [\alpha/Fe]\sim 0.1 \ dex$. The right panel of Fig.~\ref{fig:selectionandEtot}, shows the $\rm L_z$ vs $\rm E_{Tot}$ plane for the RRL stars. The rosy-brown and green points in this panel are bulge and halo/disk stars selected from \cite{Queiroz2023} based on APOGEE and Gaia DR3 surveys, and incorporated in our orbital code for consistency. The sample consisted of stars in the bulge region with limits at $\rm |X|< 5 \ kpc$,$\rm |Y|< 3.5 \ kpc$, and $\rm |Z|< 1 \ kpc$. In this panel, we identify 3 regions for the m-rich RRL stars. First, at low energy, there are 4 RRab completely inside the bulge potential. The second group at intermediate energy harbors the 3 RRL with $\rm [\alpha/Fe]\sim 0.1 \ dex$. The third region includes two stars with high energy, likely belonging to the halo. The black solid line marks $\rm L_z=0$ to separate prograde ($\rm L_z<0$) from retrograde ($\rm L_z>0$) orbits. Interestingly, m-rich stars are all in the prograde side, while m-poor and RRc stars are more distributed in both prograde and retrograde orbits. 

Fig.~\ref{fig:orbitalparams} left panel shows the actions $J_\parallel$ and $J_\perp$ of the orbits. In this plane, stars with $J_\parallel \approx -1$ are on nearly circular prograde orbits. Conversely, $J_\parallel \approx 1$ represents circular retrograde orbits. Stars on radial orbits have $J_\perp \approx -1$ while those on polar orbits have $J_\perp \approx 1$. The three m-rich stars with special symbols exhibit more circular prograde orbits compared to the rest of the stars, again indicating a probable relation with the disk. Fig.~\ref{fig:orbitalparams} middle panel shows the circularity vs the maximum height of the orbits. Two of the three m-rich stars with special symbols are located in regions associated with the disk, while the third is in a region of overlap between the disk and the bulge. In contrast, the vast majority of the stars are found in regions related to the bulge. Moreover, there are some stars related to the halo. Fig.~\ref{fig:orbitalparams} right panel shows the pericenter radius vs the eccentricity of the orbits. The three m-rich stars with special symbols show pericenter radius and eccentricity related to the disk. The majority of the stars have small values for the pericenter radius, indicating that they are very attached to the bulge component. 

Overall, we observed that not all the RRL stars present the same orbital behavior. There are RRL variables associated with the bulge, disk, and halo, with a vast majority concentrated in the bulge. Furthermore, in the group of the m-rich RRL stars, there are three with orbits that suggest they belong to the disk that also have $\rm [\alpha/Fe]\sim 0.1 \ dex$, indicating that this group, different in abundance and also in kinematics, can have a distinct origin. Hence, increasing the sample of bulge RRL with abundances and kinematics values is essential to confirm the presence of a disk population of RRL stars currently in the bulge region.

\section{Discussion and conclusion}
\label{sec:conclusions}

We present here the first spectroscopic determination of iron and $\alpha$-element abundances for RRL stars in the Galactic bulge. We analyzed GIRAFFE HR10 spectra for 60 RRab and 18 RRc stars and performed a comprehensive spectral fitting analysis using FERRE. We obtained [Fe/H] and $\rm [\alpha/Fe]$ as well as other atmospheric parameters; however, we performed a quality selection to discard values obtained with poor fitting when the absorption lines are almost lost.

The RRab stars have a median metallicity at $\rm [Fe/H]_{median}=-1.34 \pm 0.04$ dex and a $\rm [\alpha/Fe]_{median}=0.25 \pm 0.01$ dex. In the case of the RRc the values are $\rm [Fe/H]_{median}=-1.44 \pm 0.08$ dex and $\rm [\alpha/Fe]_{median}=0.24 \pm 0.03$ dex. These values are in agreement with previous results of spectroscopic abundances in RRL stars, with a slightly more m-rich value, as the bulge is more m-rich than the halo. It is important to note that our statistics are not large enough to draw strong conclusions about the mean abundances of iron and $\alpha$-elements in the bulge as a whole.  

We compared our results with those of some photometric metallicity methods from the literature. We found that several photometric methods exhibit a significant offset compared to spectroscopic measurements for RRab stars, a phenomenon observed in several studies. The origin of this offset is not completely clear. Furthermore, our data suggest a possible correlation between the metallicity difference and the [$\alpha$/Fe] ratio, which needs to be investigated further.

We investigate the potential existence of multiple populations of RRL in the bulge, as several previous studies have identified m-rich RRL ([Fe/H] > -1 dex) in our Galaxy \citep{Crestani2021,Olivares2024,DOrazi2024,Gozha2024,Prudil2025}. According to the standard, single star stellar evolution, RRL variables are m-poor stars with progenitor masses $\sim 0.6-0.8$ $M_\odot$, in the core helium-burning phase \citep{catelan15}. In order for such stars to have completed their main sequence and red giant phase, they must be as old as 10 Gyr at least. Their metallicity might reach up to solar values if they lose enough mass during the first ascent red giant phase, that is, if their $\eta$ value is higher than normal \citep[see Fig. 4 in][]{DCruz1996}. Nonetheless, several recent studies discuss the possibility that some RRL are stars formed in a binary system, with a companion with a mass in the range $\sim 0.7-2$ $M_\odot$, such that the companion strips away an important fraction of their envelope. Through this channel, the variable could be an intermediate-age star \citep{Bobrick2024}, which would explain why several RRLs are found in the MW disk \citep{Zinn2020,Matsunaga2022}. 

In a novel approach, the study of \cite{Zhang2025} shows that by means of tagging with Mira stars, there is a statistically significant portion of RRL moving as intermediate-age stars in the Galaxy (the Mira sample). There are also other very new studies showing the possibility of the existence of these stars \citep{Cabrera2024,Cuevas2024}, while others are trying to find the binary system candidates to prove the binary channel theory \citep{Abdollahi2025}. One problem for this model is the very low probability of finding an RRL star in a binary system. For instance, \cite{Kervella2019} studied the anomalies in the PMs of almost 790 RRL and found that only 7 stars are probably bound in a binary system, and the fraction of those that are in interacting binaries is significantly lower. No RRL binary system has been discovered to date to confirm this model.

The study of \cite{Gozha2024} also found m-rich RRL in the disk, but they found anomalous levels of other elements, such as sodium, aluminum, and nickel, in comparison with disk stars. They proposed an extragalactic origin for some of the RRLs. In the same way, the study of \cite{DOrazi2024} that found several m-rich stars, some of which are close to the solar metallicity, also shows that those stars are [$\alpha$/Fe] depleted. They found different abundances for other elements compared to typical disk stars, supporting the idea that disk RRL can be related to the "primordial disk" as very old fossils of that epoch.

From the kinematics and abundances, we conclude that bulge RRLs can be described as a predominantly old population with a peak in the m-poor regime, but with a wide range of metallicities, which is expected for a complex structure like the bulge. However, there are some m-rich RRL that are distinct in the $\rm [\alpha/Fe]$ ratio and with disk kinematics, and could belong to a second "younger" RRL population. This does not mean that this population is strictly young or intermediate-age. As the work of \cite{Prudil2025} suggests, there is a possibility that these two populations, both very old, may have differences of only 1 or 2 Gyr, which could account for the observed results. For instance, the oldest can be 12 Gyr, while the younger can be 10 Gyr. Therefore, RRL variables are still absolutely old populations. 

Future work will focus on increasing the statistics of RRL stars observed spectroscopically in the bulge, with the aim of proving this m-rich, [$\alpha$/Fe] depleted population by observing several other bulge fields. The future near-IR and optical spectroscopic surveys around the bulge, for instance, the 4-meter Multi-Object Spectroscopic Telescope \citep[4MOST,][]{4MOST} and the Multi-Object Optical and Near-infrared Spectrograph \cite[MOONS,][]{MOONS} surveys, will help to improve our conclusions about the chemical origin of these variable stars.

\begin{acknowledgements}

J.O.C acknowledges Marcio Catelan for the very interesting discussions and comments. Also, thanks to Andrea Kunder for her very insightful comments. Moreover, thanks to Massimo Dall'Ora for his help, advices and patience during the first visit to INAF.

J.O.C. also acknowledges support from the National Agency for Research and Development (ANID) Doctorado Nacional grant 2021-21210865, and by ESO grant SSDF21/24. 

This work is funded by ANID, Millennium Science Initiative, ICN12\_009 awarded to the Millennium Institute of Astrophysics  (M.A.S.), by the ANID BASAL Center for Astrophysics and Associated Technologies (CATA) through grant FB210003, and by  FONDECYT Regular grant No. 1230731. A. R. A. acknowledges support from DICYT through grant 062319RA. B.A.T. acknowledges support from ANID Doctorado Nacional grant 2023-21231305. M.D.L. acknowledges financial support from the project “LEGO – Reconstructing the building blocks of the Galaxy by chemical tagging” (PI: Mucciarelli) granted by the Italian MUR through contract PRIN2022LLP8TK\_001.

We gratefully acknowledge the use of data from the OGLE-IV catalog.
The OGLE project has received funding from the National Science Centre, Poland, grant MAESTRO 2014/14/A/ST9/00121 to AU.

We also acknowledge the use of data from the VVV/VVVx
ESO Public Survey program ID 179.B-2002/198.B-2004 taken
with the VISTA telescope and data products from the Cambridge
Astronomical Survey Unit (CASU). The VVV Survey data are made
public at the ESO Archive.

It also made use of NASA’s Astrophysics Data System and of the VizieR catalog access tool, CDS, Strasbourg, France \citep{simbad}.  The original description of the VizieR service was published in \citep{vizier}. 
Finally, we acknowledge the use of the following publicly available softwares: \texttt{FERRE} \citep{FERRE}, lcfit: A python package for the regression of periodic time series \citep{Dekany2019}, rr\_feh \citep{Dekany2022}, TOPCAT \citep{topcat}, pandas \citep{pandas}, IPython \citep{ipython}, numpy \citep{numpy}, matplotlib \citep{matplotlib}, Astropy, a community developed core Python package for Astronomy \citep{astropy1,astropy2} and Aladin sky atlas \citep{aladin1, aladin2}. 

\end{acknowledgements}



\bibliographystyle{aa}
\bibliography{mybiblio} 

\begin{appendix}

\section{Atmospheric parameters for RRc stars}
\label{sec:parametersRRc}

\begin{figure*}
\centering
	\includegraphics[width=12cm]{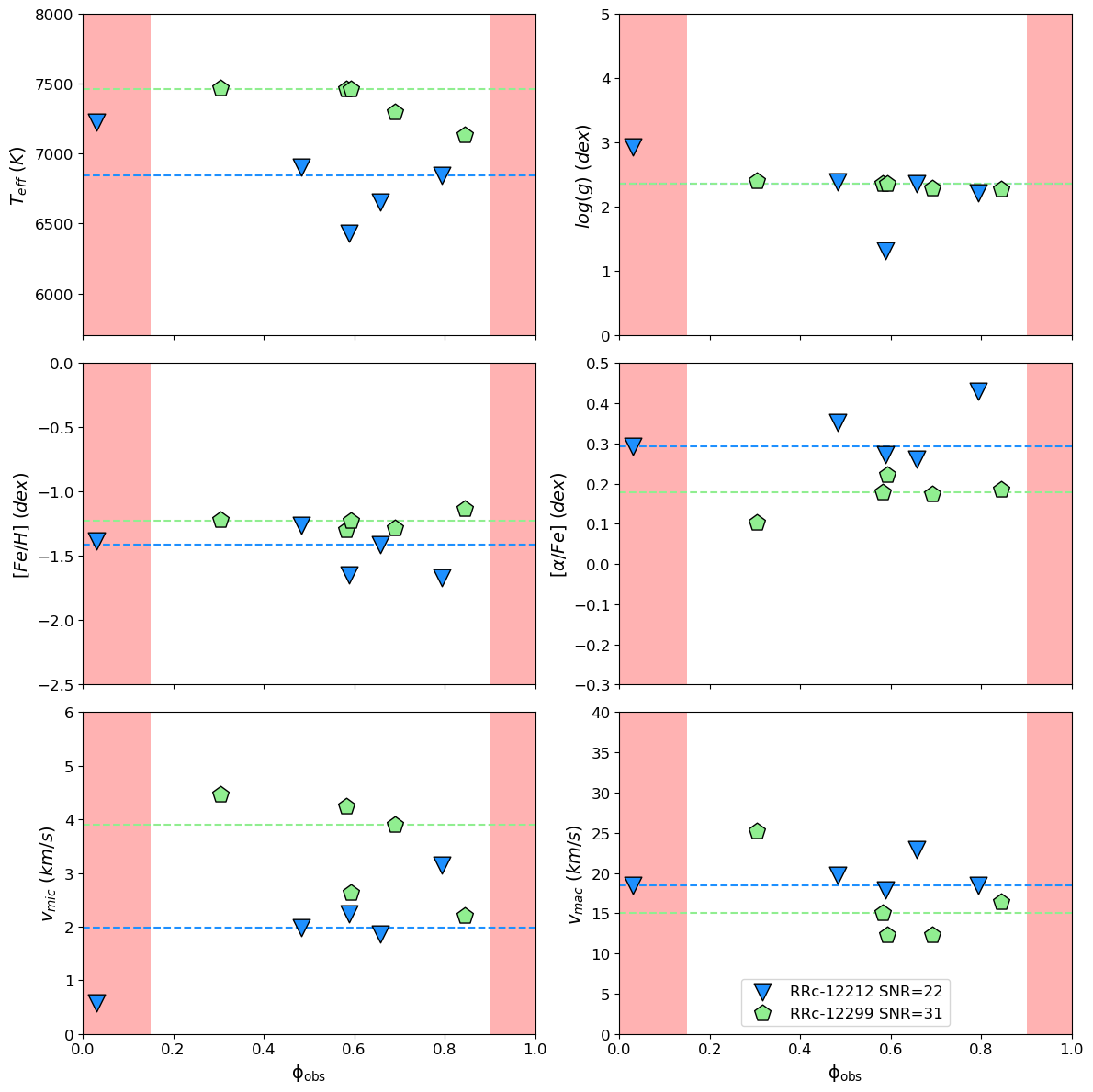}
    \caption{All the atmospheric parameter measurements per observation phase for 2 RRc stars. RRc-12212 (blue) is a m-poor star with low SNR, and RRab-12299 (green) is a m-poor star with intermediate SNR. Dashed lines indicate the mean value for the respective parameter for each star. Red regions indicate the main shock phase, where $\rm T_{eff}$ changes drastically.}
    \label{fig:6panelRRc}
\end{figure*}

We show here the variation of the atmospheric parameters for two RRc variables, with different SNR (Fig.~\ref{fig:6panelRRc}). The plot show, although the surface parameters show non negligible changes among different pulsation phases, as expected, the derived $\rm [Fe/H]$ and $\rm [\alpha$/Fe] stay consistent, with relatively little spread. This makes us confident that the abundances discussed here are robust.

\section{Errors of the atmospheric parameters for an RRL star with FERRE}
\label{sec:cornerplot}

\begin{figure*}
\centering
	\includegraphics[width=16cm]{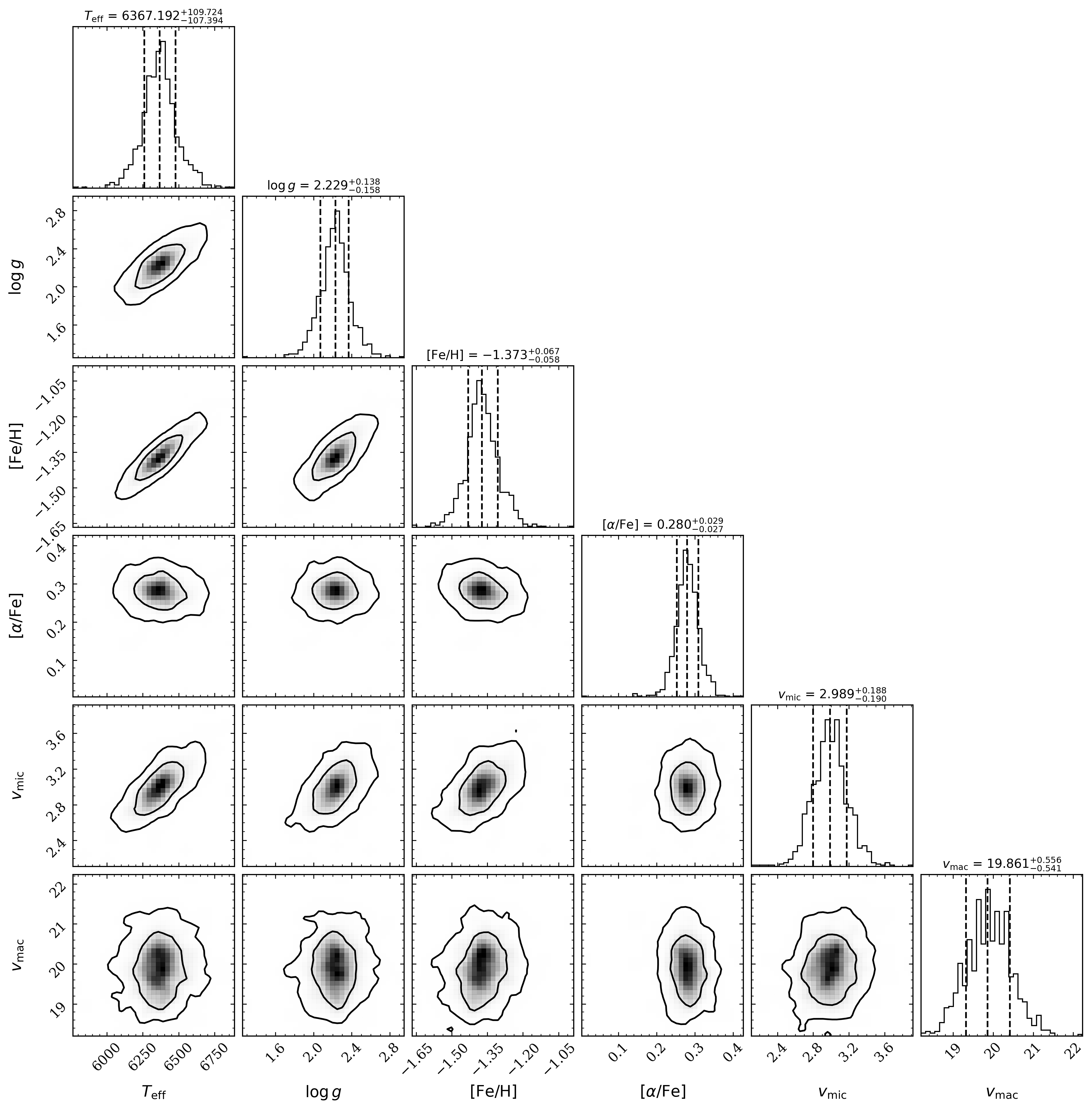}
    \caption{Corner plot with the values of the atmospheric parameters for the first observation of the star OGLE-BLG-RRLYR-11842 with SNR=33 and $\phi_{ob}=0.1$. The 1000 \textbf{data points} showed are based on an MCMC considering a Poisson error.}
    \label{fig:cornerplot}
\end{figure*}

We provide the corner plot showing the correlation between each pair of surface parameters (Fig.~\ref{fig:cornerplot}). These were obtained by perturbing each spectrum, assuming a Poisson error on its flux.

Fig.~\ref{fig:cornerplot} shows an example of a spectrum with good SNR ($\rm SNR \sim 33$) and in a good phase of observation ($\phi_{obs} = 0.43$). All atmospheric parameters follow normal distributions, and there are also some correlations between pairs of parameters. For example, for $\rm T_{eff}$ vs $\log{g}$, $\log{g}$ vs $\rm [Fe/H]$, and $\rm [Fe/H]$ vs $\rm T_{eff}$. All these relations have been previously observed in other studies \citep{Pancino2015, DOrazi2024}. More importantly, the magnitude of the errors in each independent variable is sufficiently small to ensure the validity of our results. Furthermore, for $\rm [\alpha/Fe]$, we did not observe any relation with other parameters, and the error is even smaller than for metallicity, confirming the reliability of our results. 

The results for iron and $\alpha$-element abundances are also based on the correlations present in $v_{mic}$ and the dispersion in $v_{mac}$. Adding these two extra parameters helped in reducing the uncertainties in the abundances of our interest in this work.

Another important observation is that the shapes of the correlations do not change with phase. Thus, in good conditions (good SNR and phase) the errors are small, hence the abundances are reliable.

\end{appendix}

\label{lastpage}
\end{document}